\documentclass[12pt,epsf]{article}
\usepackage{amsmath}
\usepackage{amssymb}
\usepackage{bm}
\usepackage[dvips]{graphics}

\newcommand{\be}{\begin{equation}}
\newcommand{\ee}{\end{equation}} 
\newcommand{\bea}{\begin{eqnarray}}
\newcommand{\eea}{\end{eqnarray}}
\newcommand{\nono}{\nonumber}
\newcommand{\mn}{\mu\nu}
\newcommand{\hp}{\hat{p}}
\newcommand{\hx}{\hat{x}}
\newcommand{\Bmn}{B_{\mn}}
\newcommand{\Cmn}{C^{\mn}} 
\newcommand{\Dkint}{\int\frac{d^Dk}{(2\pi)^D}}
\newcommand{\kint}{\int\frac{d^4k}{(2\pi)^4}}
\newcommand{\bphase}{-\frac{i}{2}\Cmn\frac{\partial}{\partial x^{\mu}}  
                                     \frac{\partial}{\partial
                                       y^{\nu}}} 
\newcommand{\ha}{\hat{a}} 
\newcommand{\cA}{{\cal A}} 
\newcommand{\hcA}{\hat{\cA}}

\renewcommand{\th}{\theta}
\newcommand{\bth}{\bar{\th}}
\newcommand{\sint}{\int d^4x d^2\th d^2\bth} 
\newcommand{\cint}{\int d^4x d^2\th} 
 
\newcommand{\bPhi}{\bar{\Phi}} 
\newcommand{\al}{\alpha} 
\newcommand{\ald}{\dot{\al}}

\newcommand{\bD}{\bar{D}} 
\newcommand{\spivec}[1]{\th\sigma^{#1}\bth}
\newcommand{\Phiy}{\Phi^{(y)}} 
\newcommand{\bPhiy}{\bPhi^{(y^{\dagger})}}
\newcommand{\cV}{{\cal V}}
\newcommand{\Wy}{W^{(y)}}

\newcommand{\cD}{{\cal D}}
\newcommand{\bcD}{\bar{\cD}}

\newcommand{\hPhi}{\hat{\Phi}}
\newcommand{\hbPhi}{\hat{\bPhi}}
\newcommand{\hcV}{\hat{\cV}}
\newcommand{\hW}{\hat{W}}
\newcommand{\bvol}{(2\pi)^2\sqrt{\det C}} 
\newcommand{\thint}{\int d^2\th} 
\newcommand{\bthint}{\int d^2\bth}

\newcommand{\hA}{\hat{A}}
\newcommand{\hV}{\hat{V}}
\newcommand{\ad}{\mbox{ad}\,}

\newcommand{\hLamb}{\hat{\Lambda}}
\newcommand{\bQ}{\bar{Q}}

\newcommand{\ta}{2\pi i\tau\,} 
\newcommand{\derth}[1]{\frac{\partial}{\partial\th^{#1}}}
\newcommand{\derbth}[1]{\frac{\partial}{\partial\bth^{#1}}}
\newcommand{\bxi}{\bar{\xi}}
\newcommand{\bed}{\dot{\beta}}
\newcommand{\hth}{\hat{\th}}

\newcommand{\hbth}{\hat{\bth}}
\newcommand{\hpi}{\hat{\pi}}
\newcommand{\kapint}{\int d^2\kappa} 
\newcommand{\fphase}{-\frac{1}{2}\gamma^{\al\beta}
                       \derth{\al}
                       \frac{\partial}{\partial\th^{\prime\beta}}}
\newcommand{\fvol}{\sqrt{\det \gamma}} 

\newcommand{\bpi}{\bar{\pi}}
\newcommand{\hbpi}{\hat{\bpi}}                

\newcommand{\bdel}[1]{\delta^4(\hx-{#1})}
\newcommand{\fdel}[1]{\delta^2(\hth-{#1})}
\newcommand{\vev}[1]{\left\langle{#1}\right\rangle}
\newcommand{\resol}{\frac{1}{z-\hPhi}}
\newcommand{\invvol}{\frac{i\bvol}{8\fvol}}
\newcommand{\invfac}{\frac{8\fvol}{i\bvol}}
\newcommand{\coeff}{\frac{1}{64\pi^2}}
\newcommand{\konishi}{\coeff\hW^{\al}\hW_{\al}}

\newcommand{\hN}{\hat{N}}
\newcommand{\mmfac}{\frac{\hN}{g_m}}
\newcommand{\mmfacinv}{\frac{g_m}{\hN}}
\newcommand{\mmint}{\int d^{\hN^2}\hPhi}
\newcommand{\cF}{{\cal F}}

\def\tr{\mbox{tr}}
\def\Tr{\mbox{Tr}}
\def\Str{\mbox{Str}}
\def\Stra{\Str_{x\otimes\th\otimes\bth}}
\def\Strc{\Str_{x\otimes\th}}
\def\Strac{\Str_{x\otimes\bth}}

\begin{document}
\setlength{\oddsidemargin}{0cm}
\setlength{\baselineskip}{7mm}

\begin{titlepage}  \renewcommand{\thefootnote}{\fnsymbol{footnote}}
$\mbox{ }$
\begin{flushright}
\begin{tabular}{l}
KUNS-1844 \\
hep-th/0303210\\
March 2003
\end{tabular}
\end{flushright}

~~\\
~~\\
~~\\

\vspace*{0cm}
    \begin{Large}
       \vspace{2cm}
       \begin{center}
         {Dijkgraaf-Vafa theory as large-$N$ reduction}
\\
       \end{center}
    \end{Large}

  \vspace{1cm}

\begin{center}

          Hikaru K{\sc awai}$^{ab}$\footnote
           {
e-mail address : hkawai@gauge.scphys.kyoto-u.ac.jp},
          Tsunehide K{\sc uroki}$^a$\footnote
           {
e-mail address : kuroki@gauge.scphys.kyoto-u.ac.jp}{\sc and}
          Takeshi M{\sc orita}$^a$\footnote
           {
e-mail address : takeshi@gauge.scphys.kyoto-u.ac.jp}

$^a$           {\it Department of Physics, Kyoto University,
Kyoto 606-8502, Japan}\\
$^b$           {\it Theoretical Physics Laboratory, RIKEN (The
  Institute of Physical and Chemical Research), Wako, Saitama 
351-0198, Japan}\\
\end{center}

\vfill

\begin{abstract}
\noindent 
We construct a large-$N$ twisted reduced model of 
the four-dimensional super Yang-Mills theory coupled 
to one adjoint matter. 
We first consider a non-commutative version of the 
four-dimensional superspace, and then give the mapping rule 
between matrices and functions on this space explicitly. 
The supersymmetry is realized 
as a part of the internal $U(\infty)$ 
gauge symmetry in this reduced model. 
Our reduced model can be compared with the Dijkgraaf-Vafa theory 
that claims the low-energy glueball superpotential 
of the original gauge theory is governed 
by a simple one-matrix model. 
We show that their claim can be regarded as the 
large-$N$ reduction in the sense that
the one-matrix model they proposed
can be identified with our reduced model.  
The map between matrices and functions enables us 
to make direct identities between the free energies 
and correlators of the gauge theory and the matrix model. 
As a by-product, we can give a natural explanation for 
the unconventional treatment of the one-matrix model 
in the Dijkgraaf-Vafa theory where eigenvalues lie 
around the top of the potential. 
\end{abstract}

\vfill
\end{titlepage}
\vfil\eject

\section{Introduction}
\setcounter{equation}{0}

Reduction of dynamical degrees of freedom has played a central role 
and has been paid much attention in physics. It sometimes reveals 
not only an essential structure of complicated systems, but their 
fundamental degrees of freedom. For example, 
the renormalization group \cite{Wilson}, the basic idea of which 
is the reduction of degrees of freedom 
by the block spin transformation, gives us 
insights into the universality of quantum field theory. 
Another example is the large-$N$ reduction \cite{EK,GAO}. 
This states that in the large-$N$ limit gauge theories 
in any dimensions are in a sense equivalent. Thus it can be 
regarded as a universality of the large-$N$ field theories. 
Furthermore, the reduced model brings some insights into 
the fundamental degrees of freedom of string theory. For example, 
there are a few kinds of large-$N$ reduced models 
which are conjectured to be nonperturbative formulations 
of string/M theory. One is the Matrix theory \cite{BFSS}, 
which is the large-$N$ reduced model in one dimension.  
There, the fundamental degrees of freedom are the D-particles 
whose space-time coordinates are described by large-$N$ matrices. 
Another prototype is the IIB matrix model \cite{IKKT}, 
which is the large-$N$ reduced model in zero dimension. 
Here the eigenvalues of matrices 
may be regarded as the space-time points themselves \cite{IMFA}. 

Recently Dijkgraaf and Vafa have made a claim that the exact 
low-energy superpotential for ${\cal N}=1$ gauge theories 
can be obtained by the perturbative computations in simple 
matrix models \cite{DV}. 
There, only planar diagrams of the matrix models 
contribute to the results, even if the large-$N$ limit is not 
taken in the original gauge theories. Though this claim is 
motivated by topological strings \cite{Vafa}, 
it can be proved purely by the gauge theory considerations 
in \cite{DGLVZ,CDSW}. Among others, in \cite{CDSW} a proof 
of the Dijkgraaf-Vafa theory is presented transparently 
by comparing the Schwinger-Dyson equations 
of the gauge theory and the matrix model.
 
At first sight, the Dijkgraaf-Vafa theory is another kind of 
the reduction of degrees of freedom, because it arises not from 
the large-$N$ limit, but from the supersymmetry as shown in 
\cite{CDSW}. However, in this paper, we show that the Dijkgraaf-Vafa 
theory can be regarded as the large-$N$ reduction. 
The idea is quite simple; we first consider the noncommutative 
supersymmetric gauge theory, and express it in terms of matrices. 
Here the noncommutative space-time is considered 
just as a tool to map the gauge theory to a matrix model. 
In fact, we can show that the noncommutativity does not 
contribute to the holomorphic quantities which appear in the 
Dijkgraaf-Vafa theory. Because the original gauge theory is defined 
on the superspace, we need to consider the noncommutative superspace 
where the fermionic coordinates are also noncommutative. 
As a consequence, the original gauge theory is mapped to a supermatrix 
model. We show that this model is nothing but the matrix model 
that Dijkgraaf-Vafa considered. 

This paper is organized as follows. In section \ref{SD}, 
we review the Scwinger-Dyson approach of the Dijkgraaf-Vafa theory, 
where we slightly modify the argument in \cite{CDSW}. 
In particular, we clarify the origin 
of the Konishi anomaly \cite{Konishi}, 
which plays an important role in our argument as well. 
In section \ref{basicncft}, we review 
the basic facts on the relationship between noncommutative 
gauge theories and matrix models \cite{NCYM}. 
In section \ref{susyreduced}, we construct the large-$N$ twisted 
reduced model \cite{GAO} of the noncommutative supersymmetric 
gauge theory. Then we consider the noncommutative superspace 
and the gauge theory defined on it. We show that it is mapped 
to a supermatrix model. In section \ref{DVasreduced}, we find 
a direct relation between the correlation functions and free energies 
of the supersymmetric gauge theory and the supermatrix model. 
Then we show that our supermatrix model captures the low-energy 
superpotential and incorporates the Dijkgraaf-Vafa theory. 
The point here is that 
we can make a direct map between the supersymmetric gauge theory 
and the supermatrix model.
Section \ref{discussion} is devoted to discussions. 
In appendix \ref{ncKonishi}, we give a derivation 
of the Konishi anomaly on the bosonic noncommutative space.

\section{Review of the Schwinger-Dyson approach}
\label{SD}
\setcounter{equation}{0}
We consider $\mathcal{N}=1$ $U(n)$ gauge theory 
coupled to an adjoint matter $\Phi$. 
According to the Dijkgraaf-Vafa theory, the prepotential 
of this theory is identified with the  free energy 
of a large $\hat{N}$ one-matrix model.\\

In this section, we slightly modify the proof of \cite{CDSW} 
using the Schwinger-Dyson equations. In this approach, 
the Konishi anomaly enters as a result of the regularization of 
$\delta^4(0)\delta^2(0)$, the value of the $\delta$-function 
at the origin of the superspace, that appears 
in the Schwinger-Dyson equations. 
In section 5, this quantity plays an important role 
to connect the field theory correlation functions 
with those of the matrix model.\\

The action of the $U(n)$ gauge theory is given by
\begin{align}
S =& \int d^4x d^2 \theta d^2 \bar{\theta} ~ \tr \left( e^{-V} \bar{\Phi} e^{V} \Phi \right) \nonumber \\ &+ \int d^4x 
d^2 \theta ~ \tr \left( W(\Phi) \right)
+ \int d^4x d^2 \theta ~ 2\pi i \tau ~ \tr \left( W^\alpha W_\alpha \right) + c.c. .
\label{original}
\end{align}
Here $\Phi$ is a chiral superfield in the adjoint representation of $U(n)$ , $\tau$ is the gauge coupling constant, $V
$ is the vector superfield, $W_\alpha$ is the field strength 
\begin{align}
W_\alpha = -\frac{1}{4} \bar{D} \bar{D} 
e^{-V} D_\alpha e^V, \end{align} and $W(\Phi)$ is a $(m+1)$-th order polynomial superpotential 
\begin{align}
W(\Phi)= \sum^m_{k=0}\frac{g_k}{k+1}\Phi^{k+1}.
\label{potential}
\end{align}

This theory is invariant under the translation $W_\alpha \mapsto W_\alpha - 8\pi \psi_\alpha$, where $\psi_\alpha$  is 
an anticommuting c-number, because all fields are in the adjoint representation so that the $U(1)$ gauge field is 
decoupled. Owing to this symmetry, the low energy effective action $W_{eff}$ can be expressed by a prepotential 
$\mathcal{F}$ 
\begin{align} 
W_{eff} = \int d^2 \psi~ \mathcal{F}. 
\end{align}

The $g_k$ dependence of $\mathcal{F}$ is given by the resolvent as follows. First by differentiating the partition 
function with respect to $g_k$, we obtain 
\begin{align} 
\frac{\partial}{\partial g_k}W_{eff} =\frac{\partial}{\partial g_k}\int d^2\psi~ \mathcal{F} =\left< \tr ~\frac{\Phi^
{k+1}}{k+1} \right>. \label{pre} 
\end{align} 
If we introduce the resolvent 
\begin{align}
\mathcal{R}(z)
=\frac{1}{64 \pi^2}  \tr \left( (W^\alpha - 8 \pi \psi^\alpha )( W_\alpha -8 \pi \psi_\alpha)\frac{1}{z-\Phi} \right), 
\end{align} 
the right hand side is expressed as 
\begin{align} 
\left< \tr~ \frac{\Phi^{k+1}}{k+1} \right> =\frac{1}{2 \pi i (k+1)}\int d^2\psi \oint dz \left< \mathcal{R}(z) \right> 
z^{k+1}. \label{res} 
\end{align} By comparing (\ref{pre}) and (\ref{res}), we find that the $g_k$ derivative of $\mathcal{F}$ 
can be expressed as
\begin{align}
\frac{\partial}{\partial g_k} \mathcal{F}(\mathcal{S}_i)
= \frac{1}{2 \pi i (k+1)}\oint dz~ \left< \mathcal{R}(z) \right> z^{k+1}. 
\end{align} 
We can determine the prepotential by solving the Schwinger-Dyson equations up to some ambiguities, and in order to fix 
them, we impose the following $m$ conditions 
\begin{align} 
\mathcal{S}_i = \frac{1}{2 \pi i}\oint_{C_i}dz~\left< \mathcal{R}(z) \right>, \label{b.c.1} 
\end{align} 
where $C_i$ is a contour around the $i$-th critical point. Thus we obtain  $\mathcal{F}$ as  a function of $\mathcal
{S}_i$.\\

Corresponding to the gauge theory (\ref{original}), we consider  
the $U(\hat{N})$ one-matrix model given by
\begin{align}
S_m =&\frac{\hat{N}}{g_m} \Tr~ W(\hat{\Phi}),
\label{hmm}
\end{align}
where $W$ is the same polynomial potential as (\ref{potential}) and $g_m$ is an appropriate constant of dimension 
three that makes the action dimensionless.\\

The free energy of the matrix model is defined by
\begin{align}
\exp\left( -\frac{\hat{N}^2}{g_m^2} F_m \right) &= \int d^{\hat{N}^2}\hat{\Phi}~ e^{-S_m}. \end{align}

Again the $g_k$ derivative of the free energy can be expressed by the resolvent as follows \begin{align} \frac
{\partial}{\partial g_k} F_m =& \frac{1}{2 \pi i (k+1)}\oint dz \left< R_m (z) \right> z^{k+1},\\
R_m(z) &= \frac{g_m}{\hat{N}} ~ \Tr \frac{1}{z-\hat{\Phi}} . 
\end{align} 
As we will see below, $R_m(z)$ obeys the same Schwinger-Dyson equation as $\mathcal{R}(z)$. Therefore if we impose  $m
$ conditions given by 
\begin{align}
S_i = \frac{1}{2 \pi i}\oint_{C_i}dz~ \left< R_m(z) \right>,
\label{b.c.2}
\end{align}
$ \mathcal{F}(\mathcal{S}_i)$ and $F_m(S_i)$ become identical functions up to $g_k$ independent part.\\

\subsection{Schwinger-Dyson equations of the matrix model}
In order to obtain the Schwinger-Dyson equations for $R_m$, we start from \begin{align} \int d^{\hat{N}^2}\hat{\Phi} \,\Tr 
\left( T^a \frac{1}{z-\hat{\Phi}} \right) e^{-S_m}. \end{align} By  shifting $\hat{\Phi} \mapsto \hat{\Phi} + \epsilon 
T^a$, we obtain 
\begin{align}
0=&\int d^{\hat{N}^2}\hat{\Phi} \,\Tr \left( T^a \frac{1}{z-\hat{\Phi}}T^a \frac{1}{z-\hat{\Phi}} \right) e^{-S_{m}} 
\nonumber \\ &-\frac{\hat{N}}{g_m} \int d^{\hat{N}^2}\hat{\Phi} \, \Tr \left( T^a \frac{1}{z-\hat{\Phi}} \right) \Tr \left( 
T^a W'(\hat{\Phi}) \right)e^{-S_m}. 
\end{align} 
By using the completeness of the $U(\hat{N})$ Gell-Mann matrices 
\begin{align}
\sum_a \Tr (T^a X T^a Y) &= \Tr X ~\Tr Y ,\nonumber \\
\sum_a \Tr (T^a X) \Tr (T^a Y) &= \Tr (X Y),
\end{align}
the equation becomes
\begin{align*}
0=\left< \Tr \frac{1}{z-\hat{\Phi}}~ \Tr \frac{1}{z-\hat{\Phi}} \right> 
-\frac{\hat{N}}{g_m} \left< \Tr \frac{1}{z-\hat{\Phi}} W'(\hat{\Phi}) \right>. 
\end{align*}
 Using the large $\hat{N}$ factorization, we obtain 
 \begin{align} 
 \left( \frac{g_m}{\hat{N}} \left< \Tr \frac{1}{z-\hat{\Phi}} \right> \right)^2 = \frac{g_m}{\hat{N}} \left< \Tr \left(
 \frac{1}{z-\hat{\Phi}}W'(\hat{\Phi}) 
 \right)
 \right>, \label{loop 1} 
\end{align} 
 and the right hand side can be rewritten as 
 \begin{align*}  
 \frac{g_m}{\hat{N}} ~\Tr \frac{1}{z-\hat{\Phi}} \left( W'(\hat{\Phi}) - W'(z) + W'(z) \right)  =  \frac{g_m}{\hat{N}} 
~\Tr \frac{1}{z-\hat{\Phi}} \left( W'(\hat{\Phi}) - W'(z) \right) +R_m(z)W'(z). 
 \end{align*} 
 Because the first term of the right hand side is the $(m-1)$-th polynomial, (\ref{loop 1}) can be expressed as \begin
{align} \frac{d^m}{d\,z^m } \left( R_m(z)^2  - W'(z) R_m(z) \right) =0. \label{loop 3} \end{align} This is an $m$th-
order differential equation, and as we mentioned above, we need $m$ conditions (\ref{b.c.2}) to fix the ambiguities. 
In the next subsection we show that the Schwinger-Dyson equation for $\mathcal{R}(z)$ 
in the gauge theory is identical to (\ref{loop 1}).

\subsection{Schwinger-Dyson equations of the gauge theory}
As in the matrix model, we start from
\begin{align}
\int \mathcal{D} \Phi \, \tr \left(t^a \frac{(W^\alpha(y',\theta') -8 \pi \psi^\alpha )(W_\alpha(y',\theta')- 8 \pi 
\psi_\alpha)}{z-\Phi(y',\theta')} \right) e^{-S} . 
\end{align} Again by  shifting 
\begin{align*}
\Phi(y,\theta) \mapsto \Phi(y,\theta) + \epsilon t^a \delta^4(y-y_0) \delta^2(\theta-\theta_0),
\end{align*} 
we obtain
\begin{align}
&0= \nonumber \\
&\int \mathcal{D} \Phi \, \tr\left( t^a \frac{(W^\alpha -8 \pi \psi^\alpha)(W_\alpha -8 \pi \psi_\alpha)}{z-\Phi(y',
\theta')} \delta^4(y'-y_0)\delta^2(\theta'-\theta_{0}) t^a \frac{1}{z-\Phi(y',\theta')} \right)  e^{-S}  \nonumber \\ 
&-\int \mathcal{D} \Phi \, \tr \left(t^a \frac{(W^\alpha -8 \pi \psi^\alpha)( W_\alpha -8 \pi \psi_\alpha)}{z-\Phi(y',
\theta')} \right) \tr \left( t^a W'(\Phi(y_0,\theta_0)) \right)  e^{-S}  \nonumber \\ &+\frac{1}{4}\int \mathcal{D} 
\Phi \, \tr\left(t^a \frac{(W^\alpha -8 \pi \psi^\alpha )(W_\alpha -8 \pi \psi_\alpha)}{z-\Phi(y',\theta')} \right) 
\tr \left( t^a \bar{D}^2 \bar{\Phi} (y_0,\theta_0,\bar{\theta}_0 )\right)  e^{-S} . \label{SDE} 
\end{align}
If we take the limit $(y',\theta') \mapsto (y_0,\theta_0)$, the third term becomes zero because of the property of the 
chiral ring, and there is no difficulty in the second term. However, the first term involves a singular factor 
$\delta^4(0) \delta^2(0)$, and we regularize it  by the heat kernel method as shown in appendix A:  

\begin{align}
\left.\frac{\delta \Phi^a(y_0,\theta_0 )}{\delta \Phi^b (y,\theta)}\right|_{(y,\theta)\mapsto (y_0,\theta_0)} 
=& \left.{\delta^a}_b \delta^4(y-y_0)\delta^2(\theta-\theta_0)\right|_{(y,\theta)\mapsto (y_0,\theta_0)} \nonumber \\ 
=& \frac{1}{64\pi^2}{\left( W^\alpha W_\alpha \right)^a}_b \label{anomaly}. 
\end{align}

Thus (\ref{SDE}) becomes 
\begin{align}
& \frac{1}{64 \pi^2} \left< \tr ~t^a \frac{(W^\alpha -8 \pi \psi^\alpha )(W_\alpha -8 \pi \psi_\alpha)}{z-\Phi}[W^
\beta ,[W_\beta  ,t^a]] \frac{1}{z-\Phi} \right> \nonumber \\ 
& - \left< \tr~ \frac{(W^\alpha -8 \pi \psi^\alpha )(W_\alpha -8 \pi \psi_\alpha) }{z-\Phi}W'(\Phi) \right> =0 \end
{align}

Again by the property of the chiral ring, terms containing more than two factors of $\left( W_\alpha -8 \pi \psi_
\alpha \right)$ vanish. In order to use this property, we can shift $W_\alpha \mapsto W_\alpha - 4 \pi \psi_\alpha$ in 
the Konishi anomaly, because 
such shifts of $U(1)$ part do not affect the commutator.
And by using the property of $t^a$  and the factorization of the chiral ring, we obtain 
\begin{align} 
&\left( \frac{1}{64 \pi^2} \left< \tr \frac{(W^\alpha -8 \pi \psi^\alpha)( W_\alpha -8 \pi \psi_\alpha)}{z-\Phi} 
\right> \right)^2 \nonumber \\
&= \frac{1}{64 \pi^2} \left< \tr \frac{(W^\alpha -8 \pi \psi^\alpha)(W_\alpha -8 \pi \psi_\alpha)}{z-\Phi}W'(\Phi) 
\right>. \label{loop 2} 
\end{align} 
This form is exactly the same as (\ref{loop 1}), and as in the matrix model, we can rewrite it in term of $\mathcal{R}
(z)$. 
We obtain the same differential equation as (\ref{loop 3}), and also need $m$ conditions (\ref{b.c.1}) to fix the 
ambiguities. Here we emphasize that the Konishi anomaly can be understood as a result of 
$\delta^4(0) \delta^2(0)$,
which will play a crucial role in section 5.

\section{The large-$N$ twisted reduced model}
\label{basicncft}
\setcounter{equation}{0}

In this section, we give a brief review of the large-$N$ twisted 
reduced model. 
We first introduce the noncommutative space
on which we define noncommutative field theory. 
Then we construct a mapping between field theory and matrix model. 

\subsection{Noncommutative space}
\label{bncsp}
In order to define a $D$-dimensional noncommutative space, 
we first consider a quantum mechanics of degrees of freedom $D/2$, 
which has $D/2$ momenta and $D/2$ coordinates. 
By taking appropriate linear combinations of them, we have 
operators $\hp_{\mu}$ ($\mu=1,...,D$) that satisfy 
\be
[\hp_\mu, \hp_\nu]=i\Bmn,
\label{momnc}
\ee
where $\Bmn$ is an antisymmetric tensor with real components, 
and $\mbox{rank}B=D$. Later we will see that 
(\ref{momnc}) can be obtained 
as a classical solution of a large-$N$ matrix model. 
Let $C$ be the inverse matrix of $B$ 
\be
C^{\mu\lambda}B_{\lambda\nu}={\delta^{\mu}}_{\nu},
\ee
and we define $\hx^{\mu}$ by 
\be
\hx^{\mu}=\Cmn \hp_{\nu}.
\label{defofx}
\ee
Then $\hx^{\mu}$ and $\hp_{\nu}$ satisfy the following 
commutation relations: 
\be [\hx^{\mu}, \hp_{\nu}]=i{\delta^{\mu}}_{\nu},~~~
[\hx^{\mu}, \hx^{\nu}]=-i\Cmn,~~~
[\hp_{\mu}, \hp_{\nu}]=i\Bmn.
\label{alg}
\ee

We regard $\hx^{\mu}$ ($\mu=1,...,D$) as the noncommutative coordinates 
of a $D$-dimensional noncommutative space, and consider a field theory 
defined on it. In fact, various gauge theories defined on this space 
are known to arise as the low-energy effective theory 
of string theory or M-theory \cite{SW}. In such a noncommutative field 
theory, fields or functions of $\hx^{\mu}$ have one-to-one 
correspondence to operators in the original quantum mechanics 
via the Weyl ordering,
\be
O(x)=\Dkint e^{ik_{\mu}x^{\mu}}\tilde{O}(k) ~~
\leftrightarrow~~~ \hat{O}=\Dkint e^{ik_{\mu}\hx^{\mu}}\tilde{O}(k).
\label{Weyl}
\ee
Roughly speaking, the operator $\hat{O}$ corresponding to $O(x)$ 
can be regarded as $O(\hx)$. In this correspondence, a Hermitian 
operator corresponds to a real function. {}From (\ref{Weyl}), 
we can read the following mapping rule between functions on 
the noncommutative space and operators (matrices):

\begin{enumerate}
\item If $\hat{O}_1$ and $\hat{O_2}$ correspond to $O_1(x)$ and $O_2(x)$ 
respectively, $\hat{O}_1\hat{O}_2$ corresponds to $O_1*O_2(x)$, 
where the $*$-product is defined by
\be O_1*O_2(x)=\exp\left.\left(\bphase\right)O_1(x)O_2(y)~\right|_{y=x}.
\label{*product}
\ee

\item If $\hat{O}$ corresponds to $O(x)$, 
\be
\Tr~\hat{O}=\frac{1}{(2\pi)^{D/2}\sqrt{\det C}}\int d^Dx~O(x). 
\ee 

\item If $\hat{O}$ corresponds to $O(x)$, $[\hp_{\mu}, \hat{O}]$ 
corresponds to $-i\partial_{\mu}O(x)$. 
\end{enumerate}

\subsection{Noncommutative field theory}
\label{ncft}
Now we construct a field theory defined on the noncommutative space, 
namely, noncommutative field theory. 
As the simplest example, we start with an infinite dimensional 
Hermitian matrix model 
\be
S=(2\pi)^{D/2}\sqrt{\det C}\,
\Tr\left(-\frac{1}{2}[\hat{p}_{\mu},\hat{\phi}]^2+V(\hat{\phi})\right).
\label{matrixscalar}
\ee
Here $\hat{\phi}$ and $\hp$ are Hermitian operators 
acting on a vector space, and we assume that $\hp$ form 
an irreducible representation of the algebra (\ref{momnc}). 
Using the mapping rule described above, we can reinterpret 
this theory as a real scalar field theory defined 
on the noncommutative space 
\be
S=\int d^Dx
  \left(\frac{1}{2}(\partial_{\mu}\phi)^2+V(\phi)\right)_*.
\label{ncscalar}
\ee
Here $*$ means that every product is understood as the $*$-product 
defined by (\ref{*product}). 
If we take a reducible representation of (\ref{momnc}) 
such as $\hp_{\mu}=\hp^{(0)}_{\mu}\otimes 1_n$, 
where $\hp^{(0)}_{\mu}$ is the irreducible representation 
of (\ref{momnc}), and $1_n$ is the $n\times n$ unit matrix, 
(\ref{matrixscalar}) can be mapped 
to an $n\times n$ Hermitian matrix-valued scalar field theory 
\be
S=\int d^Dx~
  \tr\left(\frac{1}{2}(\partial_{\mu}\phi)^2+V(\phi)\right)_*.
\ee

Next we turn to quantum aspects of the noncommutative field theory. 
As is well known, if we deduce the Feynman rule of (\ref{ncscalar}), 
we have the noncommutative phase factor for each vertex 
arising from the $*$-product. 
Due to this phase factor, if external momenta are much larger than 
$|B|\equiv(\sqrt{\det B})^{1/D}$, only the planar diagrams survive 
\cite{GAO}, which means that in high momentum region 
the noncommutative field theory is equivalent to the large-$N$ theory. 
On the other hand, if external momenta are much smaller than $|B|$, 
this theory is at least classically equivalent 
to the ordinary field theory on the commutative space 
because the phase factor does not contribute. 
However, in quantum theory, the noncommutative field theory has an 
effective UV cutoff of order $1/|\Cmn p_{\nu}|$ due to the phase factor, 
where $p$ is an external momentum. 
Therefore, if the theory does not have UV divergence at all 
as a field theory, we can take the low energy limit $p\rightarrow 0$ 
smoothly and the noncommutative field theory is reduced 
to the ordinary commutative field theory. 
However, if the theory has an UV divergence, 
it possibly violates this classical equivalence \cite{IRUV}.

\subsection{Noncommutative gauge theory}
\label{basicncgauge}
If we consider the gauge theory on the noncommutative space 
in the same way, we find that the corresponding matrix model 
is nothing but the large-$N$ twisted reduced model \cite{GAO}. 
In order to see this, we consider the noncommutative $U(n)$ 
gauge theory coupled to a fermion in the adjoint representation, 
\be
S=\int d^Dx
  \left(\frac{1}{g^2}\tr
  \left(-\frac{1}{4}F_{\mn}^2
        -\frac{i}{2}\bar{\psi}\Gamma^{\mu}[D_{\mu},\psi]
  \right)
  \right)_*. 
\label{ncgauge}
\ee
The corresponding matrix model is obtained via the mapping rule as 
\be
S=(2\pi)^{D/2}\sqrt{\det C}\frac{1}{g^2}\Tr
  \left(\frac{1}{4}[\hp_{\mu}+\ha_{\mu},\hp_{\nu}+\ha_{\nu}]^2
       +\frac{1}{2}\bar{\psi}\Gamma^{\mu}[\hp_{\mu}+\ha_{\mu},\psi]
  \right),
\label{matrixgauge}
\ee
up to some ambiguities coming from the ordering.
Here $\hp_{\mu}=\hp^{(0)}_{\mu}\otimes 1_n$ and the trace is taken 
over both the representation space of $\hp^{(0)}$ and $n\times n$
matrix. If we define 
\be
\hcA_{\mu}=\hp_{\mu}+\ha_{\mu},
\label{expand}
\ee
this action can be rewritten as 
\be
S=(2\pi)^{D/2}\sqrt{\det C}\frac{1}{g^2}\Tr
  \left(\frac{1}{4}[\hcA_{\mu},\hcA_{\nu}]^2
       +\frac{1}{2}\bar{\psi}\Gamma^{\mu}[\hcA_{\mu},\psi]
  \right).
\label{TEK}
\ee
As a result, $\hp_{\mu}$ dependence disappears in (\ref{TEK}). 
Instead, it has a classical solution $\hcA_{\mu}=\hp_{\mu}$ 
where $\hp$ satisfies (\ref{momnc}) 
and if we expand (\ref{TEK}) around it as (\ref{expand}), 
we recover (\ref{matrixgauge}) or, equivalently, the noncommutative 
gauge theory (\ref{ncgauge}) \cite{NCYM}. 
(\ref{TEK}) is the dimensional reduction 
of the $U(\infty)$ gauge theory with an adjoint matter 
to the zero dimension. This is nothing but the large-$N$ 
reduced model, and the expansion around the noncommutative 
background $\hcA_{\mu}=\hp_{\mu}$ is known 
as the twisted reduced model.

\section{Supersymmetric large-$N$ twisted reduced model} 
\label{susyreduced} 
\setcounter{equation}{0}

Now we construct the large-$N$ twisted reduced model of 
the supersymmetric gauge theory with an adjoint matter. 
We do this in the following two steps:

\begin{description}
\item[step1] We first describe the supersymmetric gauge theory 
on the noncommutative space in terms of superfield. 
At this stage, the four-dimensional bosonic coordinates $x^{\mu}$ 
become noncommutative, while the fermionic coordinates 
$\th^{\al}$, $\bth^{\ald}$ remain intact. 
As a result, each component of the superfield corresponds to 
a large-$N$ matrix. 

\item[step2] Next we make the fermionic coordinates  
$\th^{\al}$, $\bth^{\ald}$ noncommutative.\footnote
{Rigorously, fermionic coordinates become non-anticommutative. 
However, we call them `noncommutative' fermionic coordinates for 
simplicity.}
As a result, a superfield corresponds to a supermatrix. 
Namely, all components are encoded into a single supermatrix.    

\end{description}

\subsection{Large-$N$ reduction via superfield}
We are interested in the $U(n)$ gauge theory with one adjoint matter 
(\ref{original}). 
Before considering a noncommutative version of this theory, 
we rewrite this action in terms of fields appropriate for 
the large-$N$ reduction. When we concentrate on 
the chiral superfields as in Dijkgraaf-Vafa theory, 
convenient coordinates are given by $y^{\mu}=x^{\mu}+i\spivec{\mu}$. 
In fact, a solution of the chiral condition $\bD\Phi(x,\th,\bth)=0$  
is in general given by 
\bea
\Phi(x,\th,\bth) & = & \Phiy(x+i\spivec{},\th) \nono \\
                 & = & \exp(i\spivec{\mu}\partial_{\mu})\Phiy(x,\th)
                       \exp(-i\spivec{\mu}\partial_{\mu}),
\label{newchiralsf}
\eea
where the superscript $(y)$ indicates the representation 
in terms of $y$, $\th$ and $\bth$. 
The advantage of the $y$-representation is that a chiral superfield 
$\Phiy$ does not have $\bth$ component as above and that 
if we expand $\Phiy(y,\th)$ with respect to $\th$ as
\be
\Phiy(y,\th)=\phi(y)+\sqrt{2}\th\psi(y)+\th\th F(y), 
\label{comp} 
\ee 
all components $\phi(y)$, $\psi(y)$, $F(y)$ are independent, 
arbitrary functions of $y$. However, the natural coordinate for which 
we can introduce the noncommutativity is not $y^{\mu}$ 
but $x^{\mu}$. Therefore, we rewrite the original action 
in terms of $\Phiy(x,\th)$ appearing in (\ref{newchiralsf}). 
Similarly, we define an antichiral superfield $\bPhiy(x,\bth)$ by 
\bea
\bPhi(x,\th,\bth) 
 & = & \bPhiy(x-i\spivec{},\bth) \nono \\
 & = & \exp(-i\spivec{\mu}\partial_{\mu})\bPhiy(x,\bth)
      \exp(i\spivec{\mu}\partial_{\mu}).
\eea
Then the kinetic term of the matter field can be rewritten as 
\be
\tr(\bPhi e^V\Phi e^{-V})
=\tr(\bPhiy e^{i\spivec{\mu}\partial_{\mu}} 
        e^V e^{i\spivec{\mu}\partial_{\mu}} 
     \Phiy e^{-i\spivec{\mu}\partial_{\mu}} 
    e^{-V} e^{-i\spivec{\mu}\partial_{\mu}}).
\ee
This motivates us to define a new vector superfield
\be
e^{\cV(x,\th,\bth)}\equiv 
\exp(i\spivec{\mu}\partial_{\mu})e^{V(x,\th,\bth)}
\exp(i\spivec{\mu}\partial_{\mu}).
\label{newvectorsf}
\ee
Note that it is not a similarity transformation 
like (\ref{newchiralsf}), and $\cV(x,\th,\bth)$ is no longer 
a function but a first-order differential operator.
Obviously, $\cV^{\dagger}=\cV$. 
Thus the kinetic term becomes
\be
\tr(\bPhi e^V\Phi e^{-V})=\tr(\bPhiy e^{\cV}\Phiy e^{-\cV}). 
\ee 
Next we consider the kinetic term of the gauge field 
in (\ref{original}), which is written in terms of the field strength 
\be
W_{\al}(x,\th,\bth)
=-\frac{1}{4}\bD\bD e^{-V(x,\th,\bth)}D_{\al}e^{V(x,\th,\bth)}.
\label{fieldstrength}
\ee
It is worth noticing that this equation can be regarded as 
an equation for differential operators acting on the space 
of chiral superfields, as is the case with the field strength 
in the ordinary gauge theories. 
Namely, the action of the differential operator 
in the right-hand side of (\ref{fieldstrength}) 
on any chiral superfield is equal to the multiplication of $W_{\al}$. 
Because $W_{\al}$ is a chiral superfield, 
(\ref{newchiralsf}) tempts us to define $\Wy(x,\th)$ as 
\bea
W_{\al}(x,\th,\bth) 
 & = & \Wy_{\al}(x+i\spivec{},\th) \nono \\
 & = & \exp(i\spivec{\mu}\partial_{\mu})\Wy_{\al}(x,\th)
       \exp(-i\spivec{\mu}\partial_{\mu}).
\eea
In fact, $\Wy_{\al}$ is exactly the field strength constructed 
from $\cV$ defined in (\ref{newvectorsf}) in the same way as 
in (\ref{fieldstrength}):
\bea
\Wy_{\al} & = & e^{-A}W_{\al}e^A \nono \\
          & = & -\frac{1}{4}(e^{-A}\bD e^A)(e^{-A}\bD e^A)
                            (e^{-A} e^{-V} e^{-A})(e^{A}D_{\al} e^{-A})
                            (e^{A} e^{V} e^{A}) \nono \\
          & = & -\frac{1}{4}\bcD\bcD e^{-\cV}\cD_{\al} e^{\cV}, 
\label{newfieldstrength} 
\eea 
where $A=i\spivec{\mu}\partial_{\mu}$, and 
\bea
\cD_{\al} & = & \exp(i\spivec{\mu}\partial_{\mu})D_{\al}
                \exp(-i\spivec{\mu}\partial_{\mu})
               =\derth{\al},  \nono \\
\bcD_{\ald} & = & \exp(-i\spivec{\mu}\partial_{\mu})\bD_{\ald}
                  \exp(i\spivec{\mu}\partial_{\mu})
                 =-\derbth{\ald},
\label{newD}
\eea
are natural differential operators on the new chiral or 
antichiral superfields, $\Phiy$ and $\bPhiy$. 
Note that $\cD$ and $\bcD$ do not contain $\partial_{\mu}$ 
because their similarity transformations 
in (\ref{newD}) are inverse to each other. Note also that 
(\ref{newfieldstrength}) can again be regarded as an equation 
for differential operators acting on the space 
of the chiral superfields $\Phiy(x,\th)$. 
Using $\Wy_{\al}$, the kinetic term of the gauge field becomes 
\be
\tr(W^{\al}W_{\al})=\tr(W^{(y)\al}\Wy_{\al}).
\ee

Now we make the bosonic coordinates $x^{\mu}$ noncommutative, which 
amounts to replacing all products appearing in (\ref{original}) 
with the $*$-product defined in (\ref{*product}):
\bea
S_{NC} & = & \sint ~\left(\tr(\bPhi e^V\Phi e^{-V})\right)_* 
             \nono \\
       & + & \cint ~\ta\left(\tr(W^{\al}W_{\al})\right)_*
         +\cint ~\left(\tr ~W(\Phi)\right)_*+c.c. \nono \\
       & = &  \sint ~\left(\tr(\bPhiy e^\cV\Phiy e^{-\cV})\right)_* 
            \nono \\
       &   & +\cint ~\ta\left(\tr(W^{(y)\al}\Wy_{\al})\right)_*
             +\cint ~\left(\tr ~W(\Phiy)\right)_*+c.c.,\nono \\ 
\label{bncaction} 
\eea 
Following the prescription given in subsection \ref{bncsp}, 
we can express it in terms of matrices.
We first introduce the noncommutative space-time coordinate 
$\hx^{\mu}$ and $\hp_{\nu}$ that satisfy (\ref{alg}). 
Then by the mapping rule given in subsection \ref{bncsp}, 
we have matrix variables corresponding to the chiral superfield, 
antichiral superfield, vector superfield, and field strength, 
respectively, 
\bea
\hPhi (\th) & \leftrightarrow & \Phiy (x,\th), \nono \\
\hbPhi (\bth) & \leftrightarrow & \bPhiy (x,\bth), \nono \\ 
\hcV (\th,\bth) & \leftrightarrow & \cV (x,\th,\bth), \nono \\ 
\hW_{\al} (\th) & \leftrightarrow & \Wy_{\al} (x,\th). 
\eea
The action (\ref{bncaction}) is rewritten as $S_{red}$ 
given by 
\bea
S_{red}
 & = & \bvol\ \{\thint\bthint~\Tr(\hbPhi (\bth)\, e^{\hcV(\th,\bth)}\,
                         \hPhi (\th)\, e^{-\hcV(\th,\bth)}) \nono \\  
 & + & \thint ~\ta\Tr({\hW}^{\al}(\th)\hW_{\al}(\th))
      +\thint ~\Tr ~W(\hPhi(\th))+c.c.\},
\label{breduced}
\eea
where 
\be
\hW_{\al}=-\frac{1}{4}\bcD\bcD e^{-\hcV}\cD_{\al}e^{\hcV},
\ee
and $\Tr$ is taken over both $U(n)$ group and the representation 
space of (\ref{alg}). As seen in (\ref{TEK}), this is nothing but 
the large-$N$ twisted reduced model of the original theory 
(\ref{original}). It should be noted that $S_{NC}=S_{red}$ holds 
as an identity. 

\subsection{Properties of the supersymmetric reduced model}
In this subsection we discuss some interesting properties 
of the supersymmetric reduced model (\ref{breduced}). 

First, as is the case with the ordinary large-$N$ reduced model 
(\ref{TEK}), it does not have background dependence at all. 
In general, as we have seen in the previous section, 
$\hp_{\mu}$ appears in the action 
through the mapping rule $-i\partial_{\mu}\leftrightarrow \ad\hp_{\mu}$, 
where $\ad
\hat{O}$ denotes the adjoint action of $\hat{O}$.
However, our action does not have explicit $\hp_{\mu}$ dependence.  
In fact, the $x^{\mu}$ derivatives do not appear in the definition 
of $\cD$, $\bcD$ and $\Phiy$, as shown in (\ref{newD}) and (\ref{comp}). 
Moreover, the equation of motion of (\ref{breduced}) 
for the vector superfield $\hcV$ is given by 
\be
\cD_{\al}e^{\hcV}\hW^{\al}e^{-\hcV}=0,
\ee
which has a special solution 
\be
e^{\hcV}=e^{2\hA},
\label{back}
\ee
where $\hA=-\spivec{\mu}\hp_{\mu}$. 
As is evident from the construction in the previous subsection, 
if we expand $e^{\hcV}$ around this background as 
\be
e^{\hcV}=e^{\hA}e^{\hV'}e^{\hA}, 
\label{superexpand}
\ee
the action (\ref{breduced}) becomes 
\bea
\frac{S_{red}}{\bvol} 
 & = & \thint\bthint~\Tr(\hbPhi' e^{\hV'}\hPhi' e^{-\hV'}) \nono \\  
 & + & \thint ~\ta\Tr({\hW}^{\prime\al}\hW'_{\al})
       +\thint ~\Tr ~W(\hPhi')+c.c.,
\label{expandreduced}
\eea
where
\bea
\hPhi' & = & e^{\hA}\hPhi e^{-\hA}, \nono \\
\hbPhi' & = & e^{-\hA}\hbPhi e^{\hA}, \nono \\
\hW'_{\al} & = & e^{\hA}\hW_{\al}e^{-\hA}
             =-\frac{1}{4}\bD\bD e^{-\hV'}D_{\al}e^{\hV'}, \nono \\ 
   D_{\al} & = & e^{-\hA}\cD_{\al} e^{\hA}
             =\derth{\al}
              -(\sigma^{\mu}\bth)_{\al}\hp_{\mu}, \nono \\ 
\bD_{\ald} & = & e^{\hA}\bcD_{\ald} e^{-\hA}
             =-\derbth{\ald}
              +(\th\sigma^{\mu})_{\ald}\hp_{\mu}.
\eea
By using the mapping rule given in subsection \ref{bncsp}, 
we recover the noncommutative supersymmetric gauge theory 
(\ref{bncaction}), where $\hPhi'$, $\hbPhi'$ and $\hV'$ are mapped 
to $\Phi$, $\bPhi$ and $V$, respectively. 
This is a supersymmetric analog of what happens 
in the bosonic twisted reduced model discussed 
in subsection \ref{basicncgauge}. In particular, $\hcV$ 
in (\ref{superexpand}) is a supersymmetric analog 
of $\hcA_{\mu}$ given in (\ref{expand}). 
Indeed, it is easy to compute 
the components of $\hcV$ in (\ref{superexpand}) and to find 
that after the usual rescaling $\hV'\rightarrow 2\hV'$, 
the $\spivec{\mu}$ component of ${\hcV}$ is given 
by $-2\hcA_{\mu}=-2(\hp_{\mu}+\ha_{\mu})$, 
where $-\ha_{\mu}$ is the $\spivec{\mu}$-component of $\hV'$. 
Similarly, $\hW_{\al}$ corresponds to 
$\hat{\cal F}_{\mn}=[\hcA_{\mu},\hcA_{\nu}]$ 
in the bosonic twisted reduced model. 

Next we discuss the symmetry of the supersymmetric reduced model. 
The action (\ref{breduced}) is manifestly invariant under the following 
transformation:
\bea
\hPhi & \rightarrow & e^{-i\hLamb}\hPhi e^{i\hLamb}, \nono \\ 
\hbPhi & \rightarrow & e^{-i\hLamb^{\dagger}}\hbPhi
                       e^{i\hLamb^{\dagger}}, \nono \\ 
e^{\hcV} & \rightarrow & e^{-i\hLamb^{\dagger}}e^{\hcV}e^{i\hLamb},
\label{sym}
\eea 
where $\hLamb$ is an arbitrary chiral superfield, $\bcD\hLamb=0$. 
This symmetry is the counterpart of the ordinary gauge symmetry 
of the supersymmetric gauge theory (\ref{original}). 
Remarkably, this symmetry includes the supersymmetry 
of the corresponding noncommutative gauge theory (\ref{bncaction}). 
In this sense, in the twisted reduced model (\ref{breduced}), 
the gauge symmetry and the supersymmetry are unified. 
This fact can be shown as follows: take the background (\ref{back}) 
and make the expansion around it as (\ref{superexpand}), 
then we get the action (\ref{expandreduced}). In terms of the fields 
appearing in (\ref{expandreduced}), the gauge transformation becomes 
\bea 
\hPhi' & \rightarrow & e^{-i\hLamb'}\hPhi' e^{i\hLamb'}, \nono \\ 
\hbPhi' & \rightarrow & e^{-i\hLamb^{\prime\dagger}}\hbPhi'  
                        e^{i\hLamb^{\prime\dagger}}, \nono \\ 
e^{\hV'} & \rightarrow & e^{-i\hLamb^{\prime\dagger}}e^{\hV'}
                         e^{i\hLamb'}, 
\label{expandsym}
\eea
where $e^{\hLamb'}=e^{\hA}e^{\hLamb}e^{-\hA}$. Note that if 
$\hLamb$ is chiral, namely, $\bcD_{\ald}\hLamb=0$, then 
$\hLamb'$ is chiral, namely, $\bD_{\ald}\hLamb'=0$.
Now we consider a particular gauge transformation (\ref{sym}) 
with $\hLamb$ given by 
\be
\hLamb=\xi^{\al}\derth{\al}
      +\bxi^{\ald}\left(-\derbth{\ald}
                        -2(\th\sigma^{\mu})_{\ald}\hp_{\mu}\right).
\ee
If we expand the theory around the background (\ref{back}), 
this symmetry becomes the gauge symmetry (\ref{expandsym}) 
with $\hLamb'$ given by 
\be
\hLamb'=\xi^{\al}\left( \derth{\al}
                       +(\sigma^{\mu}\bth)_{\al}\hp_{\mu}\right)
      +\bxi^{\ald}\left(-\derbth{\ald}
                        -(\th\sigma^{\mu})_{\ald}\hp_{\mu}\right)+\lambda,
\ee
where $\lambda$ is a complex number. 
Because $\hLamb^{\prime\dagger}=\hLamb'$, the infinitesimal form 
of the gauge transformation (\ref{expandsym}) is given by 
\bea
\delta\hPhi' & = & \ad(-i\hLamb')\hPhi'
               =(-i\xi^{\al}Q_{\al}-i\bxi^{\ald}\bQ_{\ald})\hPhi',
               \nono \\ 
\delta\hbPhi' & = & \ad(-i\hLamb^{\prime\dagger})\hbPhi'
                =(-i\xi^{\al}Q_{\al}-i\bxi^{\ald}\bQ_{\ald})\hbPhi',
                \nono \\ 
\delta V' & = & \ad(-i\hLamb')V'
            =(-i\xi^{\al}Q_{\al}-i\bxi^{\ald}\bQ_{\ald})V',  
\eea        
where 
\bea
Q_{\al} & = & \derth{\al}+(\sigma^{\mu}\bth)_{\al}\ad\hp_{\mu}, 
                                                         \nono \\ 
\bQ_{\ald} & = & -\derbth{\ald}-(\th\sigma^{\mu})_{\ald}\ad\hp_{\mu}.
\label{ncsupercharge}
\eea
Note that the transformation law for $V'$ becomes a similarity 
transformation due to the Hermiticity of $\hLamb'$. 
(\ref{ncsupercharge}) are equivalent to the ordinary supercharges 
in the noncommutative gauge theory (\ref{bncaction}) 
via the mapping rule $\ad\hp_{\mu}\leftrightarrow -i\partial_{\mu}$. 
Therefore, we have shown that once we expand the original model 
(\ref{breduced}) around the background (\ref{back}), we get the 
noncommutative gauge theory (\ref{bncaction}) and its 
supersymmetry originates from the gauge symmetry (\ref{sym}) 
of the original model. In the ordinary field theory, 
what makes difference between 
the gauge symmetry and the supersymmetry is that 
the former is generated by functions of $x^{\mu}$,
while the latter by the derivative $\partial/\partial x^{\mu}$, 
$\partial/\partial\th^{\al}$ and $\partial/\partial\bth^{\ald}$. 
However, in the large-$N$ twisted reduced model, 
or in the noncommutative space,
there is no definite 
difference between the `coordinate' and the `momentum' as we can see 
from eq.(\ref{defofx}). This is the reason why the gauge symmetry 
and the supersymmetry are unified in (\ref{breduced}).

\subsection{Noncommutative superspace and supermatrix model} 
\label{ncsusp} 
As mentioned in the beginning of this section, the next task is to 
introduce the noncommutative fermionic coordinates as well as the 
bosonic coordinates. Then it is expected that a field depending 
on the noncommutative fermionic coordinates 
$\th$ or $\bth$ is also mapped to a matrix, 
as a field on the noncommutative bosonic coordinates 
$\hx^{\mu}$ becomes the large-$N$ matrix. It is shown that a field 
on the noncommutative superspace is described by a supermatrix. 

We begin with introducing a noncommutativity into the fermionic 
coordinates as 
\be
\{\hth^{\al},\hth^{\beta}\}=\gamma^{\al\beta},~~~
\{\hbth^{\ald},\hbth^{\bed}\}=\gamma^{*\,\ald\bed},
\label{fnc}
\ee
where $\gamma^{\al\beta}$ is a symmetric matrix.
In what follows, we consider only $\hth^{\al}$ part because 
$\hbth$ can be treated in the same way by replacing
$\gamma^{\al\beta}$ with $\gamma^{*\,\ald\bed}$. 
By using the $SL(2,{\mathbf C})$ transformation, 
$\gamma^{\al\beta}$ can be taken in the following form 
without loss of generality:
\be
(\gamma^{\al\beta})=\left(\begin{array}{cc} \gamma & 0 \\
                                  0   & \gamma 
             \end{array}
       \right),~~~\gamma\in{\mathbf C}.
\ee
In this case, $\hth^{\al}$ can be represented in terms of 
Pauli matrices as 
\be 
\hth^1=\sqrt{\gamma}\sigma^1,~~~\hth^2=\sqrt{\gamma}\sigma^2.
\label{standard}
\ee
Let $\beta$ be the inverse matrix of $\gamma$ 
\be \gamma^{\al\gamma}\beta_{\gamma\beta}={\delta^{\al}}_{\beta},
\ee
and define $\hpi_{\al}$ by 
\be
\hpi_{\al}=\beta_{\al\beta}\hth^{\beta}.
\ee
Then $\hth^{\al}$ and $\hpi_{\beta}$ satisfy the following 
anticommutation relations:
\be
\{\hth^{\al},\hpi_{\beta}\}={\delta^{\al}}_{\beta},~~~
\{\hth^{\al},\hth^{\beta}\}=\gamma^{\al\beta},~~~
\{\hpi_{\al},\hpi_{\beta}\}=\beta_{\al\beta}.
\ee
As in the case of the bosonic noncommutative space, 
we regard $\hth^{\al}$ as the noncommutative fermionic coordinates 
and make a correspondence between a function on this space 
and an operator (a matrix) via the Weyl ordering: 
\be
O(\th)=\kapint ~e^{i\th^{\al}\kappa_{\al}}\tilde{O}(\kappa)
~~\leftrightarrow~~~
\hat{O}=\kapint ~e^{i\hth^{\al}\kappa_{\al}}\tilde{O}(\kappa).
\label{fWeyl}
\ee
As before, the operator $\hat{O}$ is nothing but the Weyl ordered 
form of $O(\hth)$. 

It is interesting to consider what corresponds to the fermionic 
integration $\thint$ in the space of operators 
under the correspondence (\ref{fWeyl}). 
In general, a function of $\th$ can be expanded as 
\bea
\Phi(\th) & = & \phi+\sqrt{2}\th^{\al}\psi_{\al}+\th\th F \nono \\
          & = & \phi+\sqrt{2}\th^{\al}\psi_{\al}-2\th^1\th^2 F. \eea 
then $\thint \Phi(\th)=F$. On the other hand, 
the operator corresponding to $\Phi(\th)$ is given by its Weyl 
ordered form
\bea
\Phi(\hth) & =& \phi+\sqrt{2}\hth^{\al}\psi_{\al}
                -(\hth^1\hth^2-\hth^2\hth^1)F \nono \\
            & =& \phi+\sqrt{2}\hth^{\al}\psi_{\al}+\hth\hth F. 
\label{superphi} 
\eea 
Because we have fixed the representation of $\hth$ as (\ref{standard}), 
$\hth^1\hth^2-\hth^2\hth^1=2i\gamma\sigma^3$ and therefore, 
if we define a $\Str_{\th}$ as 
\be
\Str_{\th}(\hPhi) \equiv 2\Tr(\sigma^3\hPhi),
\ee
then
\be
\Str_{\th}(\hPhi)=-8i\gamma F=-8i\gamma\thint \Phi(\th).
\ee
Thus as in the case of $\hx^{\mu}$, it is easy to derive 
the following mapping rule from (\ref{fWeyl}):

\begin{enumerate}
\item If $\hat{O}_1$ and $\hat{O_2}$ correspond to 
$O_1(\th)$ and $O_2(\th)$ respectively, $\hat{O}_1\hat{O}_2$ 
corresponds to $O_1\star O_2(\th)$, 
where the fermionic $\star$-product is defined by,
\be
O_1\star O_2(\th) 
=\exp\left.\left(\fphase\right)O_1(\th)O_2(\th')~\right|_{\th'=\th}.
\label{f*product}
\ee

\item If $\hat{O}$ corresponds to $O(\th)$, 
\be
\Str_{\th}(\hat{O})=-8i\fvol\thint~O(\th).
\ee
\item If $\hat{O}$ corresponds to $O(\th)$, $[\hpi_{\al},\hat{O}\}$ 
corresponds to $\partial/\partial\th^{\al}O(\th)$, 
where the commutator or anticommutator is taken 
according to the statistics of $\hat{O}$.  
\end{enumerate}

Now we define the large-$N$ twisted reduced model
on the noncommutative superspace.
First we replace the product in (\ref{breduced}) with the 
$\star$-product in the space of $\hat{\theta}$ 
and $\hat{\bar{\theta}}$ defined above. 
We then rewrite the action using the mapping rule given above, 
and obtain 
\bea 
S_{smm} & = & \frac{i^2 (2\pi)^2 \sqrt{\det C}}
                   {8^2\sqrt{\det\gamma}\sqrt{\det\gamma^*}}
              \ \Stra (\hbPhi e^{\hcV}\hPhi e^{-\hcV}) \nono \\
        & + & \frac{i (2\pi)^2 \sqrt{\det C}}{8\sqrt{\det\gamma}}
              \ \{\ta\Strc ({\hW}^{\al}\hW_{\al})
             +\Strc (W(\hPhi))\}+c.c.,
\label{supermatrix}
\eea
where
\be
\hW_{\al}
=-\frac{1}{4}\ad\hbpi_{\ald}\ad\hbpi^{\ald} 
  e^{-\hcV}\ad\hpi_{\al}e^{\hcV}, 
\ee 
and $\ad\hpi_{\al}$ is defined by 
\be
\ad\hpi_{\al}\hat{O}
=\left\{\begin{array}{cl}
             [\hpi_{\al},O] & \mbox{for even }\hat{O}, \\
            \{\hpi_{\al},O\} & \mbox{for odd }\hat{O},
        \end{array} 
 \right.
\ee
and $\ad\hbpi_{\ald}$ is similarly defined. 
$\Stra$ means taking trace in the bosonic space 
of $\hx^{\mu}$ and supertraces in the fermionic spaces of 
$\hat{\theta}$ and $\hat{\bar{\theta}}$.
Here, as usual in the large-$N$ reduced model, 
the $U(n)$ gauge group and the bosonic noncommutative space 
are unified.
Similarly, $\Strc$ and $\Strac$ can be defined unambiguously.\footnote
{It is likely that by expanding (\ref{supermatrix}) 
around a classical solution such as (\ref{back}), 
we can obtain a field theory on the noncommutative superspace 
where every product is defined by the combination 
of the bosonic $*$ and fermionic $\star$ product. 
However, it does not seem straightforward to generalize 
the classical solution (\ref{back}) to $\gamma\neq 0$ case. 
Moreover, it is easy to find that if we expand (\ref{supermatrix}) 
around (\ref{back}), (\ref{supermatrix}) is not simply reduced to 
the ordinary noncommutative gauge theory because there appear 
terms which cannot be interpreted as a local field 
in the noncommutative field theory. Of course in the limit 
$\gamma,\gamma^*\rightarrow 0$ this theory is reduced to 
the noncommutative gauge theory (\ref{bncaction}).}  
In the supermatrix model (\ref{supermatrix}), 
the chiral or antichiral condition becomes 
\be
\ad\hbpi_{\ald}\hat{O}=0,~~~\ad\hpi_{\al}\hat{O}=0,
\ee
which indicate that $\hat{O}$ does not have 
$\hbth$ dependence or $\hth$ dependence, respectively. 
It is evident by the mapping rule that $\hPhi$ and $\hW_{\al}$ 
in (\ref{supermatrix}) are chiral supermatrices, 
while $\hbPhi$ is an antichiral supermatrix. 
It is also obvious by construction that in the fermionic commutative 
limit $\gamma, \gamma^*\rightarrow 0$, supermatrices 
in (\ref{supermatrix}) tend to corresponding fields in
(\ref{bncaction}) as follows:
\bea
\hPhi & \rightarrow \Phiy (x,\th), \nono \\
\hbPhi & \rightarrow \bPhiy (x,\bth), \nono \\
\hcV & \rightarrow \cV (x,\th,\bth), \nono \\
\hW_{\al }& \rightarrow \Wy_{\al}(x,\th).
\eea

\section{Dijkgraaf-Vafa theory as the large-$N$ reduction} 
\label{DVasreduced} 
\setcounter{equation}{0} 
In this section we show that 
the Dijkgraaf-Vafa theory can be understood in terms 
of the large-$N$ reduced model. 

To begin with, we note that the holomorphic quantities 
we have discussed in section \ref{SD} in the original gauge theory 
are not affected by the bosonic noncommutativity $C^{\mn}$. 
These quantities carry zero external momenta, and do not have 
UV divergences. Therefore we expect that they do not depend 
on the bosonic noncommutativity
$C^{\mn}$ for the reason explained in subsection \ref{ncft}.
In fact, as shown in \cite{DGLVZ,CDSW}, in the perturbative 
expansion, only the planar diagrams contribute 
to them\footnote
{This fact is a consequence of the chiral ring \cite{CDSW}. 
It is easy to check that this structure persists in the bosonic 
noncommutative gauge theory (\ref{bncaction}).}. 
It indicates that they have no dependence on $C^{\mn}$,
because the noncommutative phase factors 
cancel in planar diagrams \cite{GAO}. 
Therefore, as far as the holomorphic quantities which appears 
in the Dijkgraaf-Vafa theory are concerned, 
the same results can be obtained, even if we use the noncommutative 
version of the original theory (\ref{bncaction}) 
or equivalently, its large-$N$ reduced model (\ref{breduced}). 
This further implies that we can compute them 
via the supermatrix model (\ref{supermatrix}), 
if we take the commutative limit $\gamma\rightarrow 0$, 
$\gamma^*\rightarrow 0$ of the fermionic coordinates.  
In this section we discuss how to do this.

\subsection{Equivalence of the correlation function}
In order to express the correlation functions in (\ref{original}) 
in terms of the supermatrix model (\ref{supermatrix}), 
we use the following simple but important equations:
\bea
\bdel{x}^2 & = & \frac{1}{\pi^4\det C}, \nono \\
\fdel{\th}^2 & = & -4\det\gamma.
\label{delta^2}
\eea
The proof is straightforward, if we use the definitions 
\bea
\bdel{x} & = & \kint e^{ik_{\mu}(\hx^{\mu}-x^{\mu})}, \nono \\ 
\fdel{\th} & = & 4\kapint~e^{i(\hth^{\al}-\th^{\al})\kappa_{\al}},
\eea
and take $\lim_{y\rightarrow x}\bdel{x}\bdel{y}$ and 
$\lim_{\th'\rightarrow\th}\fdel{\th}\fdel{\th'}$.
In the commutative limit $C\rightarrow 0$ 
of the bosonic coordinates,
the usual result in the bosonic commutative space 
$\delta^4(0)=\infty$ is reproduced:
\be
\delta^4(0)\bdel{x}=\bdel{x}^2\rightarrow\infty. 
\ee
Similarly, in the commutative limit $\gamma\rightarrow 0$
of the fermionic coordinates, we have 
\be 
\delta^2(0)\fdel{\th}=\fdel{\th}^2\rightarrow 0,
\ee
which is the usual result in the commutative fermionic space. 
Eqs.(\ref{delta^2}) are quite peculiar 
to the noncommutative space which is essentially regularized 
by the noncommutativity and gives the finite result in nature.  
{}From (\ref{delta^2}), we can derive an identity 
\be
\left(\invvol\bdel{x}\fdel{\th}\right)^2=1.
\label{identity}
\ee
On the other hand, if a chiral superfield $O^{(y)}(x,\th)$ 
in the bosonic noncommutative gauge theory  (\ref{bncaction}) 
corresponds to a chiral supermatrix $\hat{O}$ 
in the supermatrix model (\ref{supermatrix}) 
in the $\gamma\rightarrow 0$ limit, 
we obtain by the mapping rule 
\bea
\lefteqn{\invvol\Strc(\hat{O}\bdel{x}\fdel{\th})} \nono \\
 & & \rightarrow \int d^4x' d^2\th' ~\tr\,(O^{(y)}(x',\th')
       \delta^4(x'-x)\delta^2(\th'-\th)) \nono \\
 & & = \tr~(O^{(y)}(x,\th)),~~~\mbox{as}~\gamma\rightarrow 0, 
\label{evaluation} 
\eea 
where the trace is taken over the $U(n)$ group. Therefore, 
in $\gamma\rightarrow 0$ limit, the operator 
in the left-hand side corresponds to the local field 
in (\ref{bncaction}). Namely, the action 
$\Strc(\bdel{x}\fdel{\th}~\cdot)$ on a supermatrix essentially 
evaluates the corresponding field at $x$, $\th$ 
in the noncommutative field theory side. 

In the supermatrix model (\ref{supermatrix}), 
a fundamental correlator is the resolvent, 
\be
\vev{\Strc\frac{1}{z-\hPhi}}.
\label{resolvent}
\ee
Using (\ref{identity}) and (\ref{evaluation}), 
we can find what kind of field in (\ref{bncaction}) 
corresponds to (\ref{resolvent}) in the $\gamma\rightarrow 0$ limit 
as follows:  
\bea
\lefteqn{\invfac\Strc\left(\resol\right)}
\nono \\
 & & =\invfac\left(\invvol\right)^2
      \Strc\left(\resol\bdel{x}^2\fdel{\th}^2\right) \nono \\
 & & =\invvol\Strc\left(\resol\delta^4(0)\delta^2(0)\bdel{x}\fdel{\th}
               \right) \nono \\
 & & \rightarrow\coeff\tr\left(\frac{W^{(y)\al}(x,\th)\Wy_{\al}(x,\th)}
                          {z-\Phiy(x,\th)}
               \right)_*,~~~\mbox{as}~\gamma\rightarrow 0 
\label{derivation} 
\eea 
where we have used the Konishi anomaly \cite{Konishi} 
in the bosonic noncommutative space 
\be
\lim_{\gamma\rightarrow 0}\delta^4(0)\delta^2(0)=\konishi.
\label{anomaly}
\ee
In appendix \ref{ncKonishi}, we give a derivation of this equation. 
Because $\lim_{\gamma\rightarrow 0}S_{smm}=S_{red}=S_{NC}$, 
we thus conclude 
\be
\lim_{\gamma\rightarrow 0} \invfac\vev{\Strc\left(\resol\right)}
=\coeff\vev{\tr\left(\frac{W^{(y)\al}(x,\th)\Wy_{\al}(x,\th)}
                          {z-\Phiy(x,\th)}
              \right)_*}_{NC}
\label{correlatorrel}
\ee
where the subscript $NC$ indicates the correlation function 
in the theory with the bosonic noncommutativity (\ref{bncaction}). 
 
{}From the point of view of the supermatrix model (\ref{supermatrix}), 
holomorphic quantities such as (\ref{resolvent}) are determined 
by the holomorphic part of the action. In particular, 
they do not depend on the kinetic term of the chiral superfield 
$\hPhi$ (the first term) in (\ref{supermatrix}) 
and we can neglect it in the computation of (\ref{resolvent}). 
Once we do it, it is evident that 
the kinetic term of the vector superfield $\hcV$ (the second term) 
can be also neglected because $\hPhi$ and $\hcV$ are now decoupled. 
Thus the holomorphic potential term 
\be
S_{smm}^{hol}=\invvol\Strc(W(\hPhi)),
\label{DV}
\ee
is only the relevant term to (\ref{resolvent}). 
This fact can be explicitly checked if we consider 
the Schwinger-Dyson equation for (\ref{resolvent}) in
(\ref{supermatrix}) 
where the kinetic terms of $\hPhi$ and $\hcV$ do not play any roles. 
Thus as far as (\ref{resolvent}) is concerned, we can further 
reduce the action from (\ref{supermatrix}) to (\ref{DV}). 

Here we make a remark about a relation between $\sqrt{\det C}$ 
and the rank of the supermatrix. Suppose we represent the Heisenberg 
algebra (\ref{momnc}) by the $N\times N$ matrix, 
where we take the large-$N$ limit at the end. Then the matrices in the 
twisted reduced model has rank $\hN=nN$. Of course, as we have seen 
in subsection \ref{basicncgauge}, there is no notion of $n$ and $N$ 
in the twisted reduced model itself. 
It is the background $\hp_{\mu}=\hp^{(0)}_{\mu}\otimes 1_n$ 
that brings the notion of the rank of the gauge group $n$ and 
that of the noncommutative space $N$ in the model.   
As is well known, from the point of view of the twisted reduced model, 
$\det C$ is proportional to $\hN$ as
\be
\sqrt{\det C}=\frac{\hN}{(2\pi)^{\frac{D}{2}}\Lambda^D}.
\label{mmvol}
\ee  
This can be seen by considering the minimal twist configuration for 
$\hp_{\mu}$, which is the basic classical solution 
in the twisted reduced model and satisfies 
\be
e^{ia\hp_{\mu}^{(i)}}e^{ia\hp_{\nu}^{(i)}}
=e^{ia\hp_{\nu}^{(i)}}e^{ia\hp_{\mu}^{(i)}}e^{-i\frac{2\pi}{\hN^{(i)}}}.
\ee
Here $a=1/\Lambda$ is the lattice spacing,  $i$ ($i=1,...,D/2$) 
is the label of the pair of the direction subject to the twist, 
and $\hN_i$ is the rank of the matrix $\hp_{\mu}^{(i)}$. 
Therefore we have 
\be
a^2B_{\mn}=\frac{2\pi}{\hN_i}, 
\ee
which leads to (\ref{mmvol}) by using $\hN=\Pi_{i=1}^{D/2}\hN_i$. 
Eq.(\ref{mmvol}) can also be understood as follows. 
In the reduced model, we first fix a UV cutoff $\Lambda$. 
A matrix with rank $\hN$ describes $\hN$ degrees of freedom 
because $\tr~1_{\hN}=\hN$. Each degree of freedom has the mass 
dimension 1 as seen from (\ref{expand}) and 
has a volume $\sim\sqrt{\det B}$ in the momentum space 
due to (\ref{momnc}), which effectively gives the IR cutoff. 
Thus we get 
\be
\Lambda^D\sim\hN\sqrt{\det B}, 
\ee
which is consistent with (\ref{mmvol}). In the large-$\hN$ limit, 
the volume of each degree of freedom in the momentum space 
becomes small, and therefore the IR cutoff 
in the momentum space tends to zero. 
This agrees with the remark we made in subsection \ref{ncft} 
that it is the high energy region much larger than $|B|$ 
that the description by the large-$N$ field theory becomes good. 

On the other hand, as we have explained in subsection \ref{ncft}, 
in the low energy region much smaller than $|B|$, 
the description via the noncommutative field theory becomes good 
in the sense that it is well approximated by its commutative limit. 
In this case, the noncommutativity 
in the coordinate space brings an effective UV cutoff, 
and it is convenient to consider in the coordinate space. 
In order to go to the description by the noncommutative field theory, 
we have taken the background $\hp_{\mu}=\hp^{(0)}_{\mu}\otimes 1_n$, 
and expanded the theory around it. 
Then our space-time consists of $N$ (not $\hN$) unit cells
of volume $\sim\sqrt{\det C}$. 
Therefore the total volume $V$ is given by 
\be 
V\sim \frac{\hN\sqrt{\det C}}{n}. 
\label{ncvol}
\ee   

Turning back to our model (\ref{DV}), this observation 
leads us to define
\be
\invvol=\mmfac,
\label{gs}
\ee
where we have introduced a formal parameter $g_m$ 
with the mass dimension 3 on the dimensional grounds, 
and have used (\ref{mmvol}) because 
we are now at the standpoint of the matrix model. Various factors 
such as $\sqrt{\det\gamma}$ have been absorbed 
in the definition of $g_m$, and (\ref{DV}) becomes 
\be
S_{smm}^{hol}=\mmfac\Strc(W(\hPhi)). 
\label{onesupermm}
\ee
We can start from this action, 
and compute the $\gamma\rightarrow 0$ limit 
of the resolvent 
\be
\mmfacinv\vev{\Strc\left(\resol\right)}.
\label{mmresolvent}
\ee
As a matter of fact, the $g_m$ 
dependence disappears, if we express the resolvent in terms of 
$S_i$'s constructed from (\ref{mmresolvent}), 
which indicates that the result has no explicit dependence on 
$C$, $\gamma$ and $\gamma^*$. It can be checked directly 
by considering the Schwinger-Dyson equation for (\ref{mmresolvent}) 
in the one-supermatrix model (\ref{onesupermm}). 
Therefore, we can take the commutative limit $C\rightarrow 0$ 
of (\ref{correlatorrel}) to obtain 
\be
\mmfacinv\vev{\Strc\left(\resol\right)}
=\coeff\vev{\tr\left(\frac{W^{\al}(x,\th)W_{\al}(x,\th)}
                          {z-\Phi(x,\th)}
              \right)},
\label{conclusion}
\ee
where the correlation function in the right-hand side is 
the one in the original gauge theory (\ref{original}). 
This argument supports the observation given at the beginning 
of this section. There, we have noted that the holomorphic quantities 
without UV divergence in the Dijkgraaf-Vafa theory are not influenced 
by the bosonic noncommutativity $C^{\mn}$ from the point of view 
given in subsection \ref{ncft} or, more explicitly, from that of 
the perturbation theory. 
Thus we establish the equivalence 
between the resolvent (\ref{mmresolvent}) of the one-supermatrix model 
(\ref{onesupermm}) and the correlation function 
in the right-hand side of eq.(\ref{conclusion}) 
in the supersymmetric gauge theory (\ref{original}). 
This is nothing but the Dijkgraaf-Vafa theory, except that 
we should consider the one-supermatrix model 
rather than the ordinary Hermitian one-matrix model. 
Later we will discuss this point in more detail. 
In fact, it is pointed out in \cite{DV2} 
that the effective superpotential of the gauge theory 
can be computed by a supermatrix model.  
Note that we have seen the Dijkgraaf-Vafa theory by constructing 
a direct mapping (\ref{conclusion}) 
between the correlators of the gauge theory and the supermatrix model, 
instead of comparing the formal structures 
of the Schwinger-Dyson equation.

\subsection{Equivalence of the free energy}
\label{equivfreeE}
In this section we show that 
in the limit $\gamma, \gamma^* \rightarrow 0$, 
the free energy of the supermatrix model (\ref{onesupermm}) 
becomes the prepotential of the original gauge theory 
(\ref{original}). 

We define the free energy of the supermatrix model (\ref{onesupermm}) by 
\be
\exp\left(-\frac{\hN^2}{g_m^2}F_{m}\right)
=\mmint \exp\left(-\mmfac\Strc(W(\hPhi))\right).
\label{mmfreeE}
\ee
It is easy to check that $F_m$ is equal to
the the holomorphic part of the free energy $F_{smm}$ 
of the large-$N$ reduced model (\ref{supermatrix})
\bea
\lefteqn{\exp\left(-\frac{\hN^2}{g_m^2}F_{smm}\right)} \nono \\  
& = & \mmint\int d^{\hN^2}\hV 
       \exp\left(-\mmfac\{ \ta\Strc(\hW^{\al}\hW_{\al})
                          +\Strc(W(\hPhi))\}\right). 
\label{supermmfreeE} 
\eea 
Here we have omitted the kinetic term and the anti-holomorphic term
from (\ref{supermatrix}), because they do not contribute to the
holomorphic part of the free energy due to the holomorphy. 
Then $\hPhi$ and $\hcV$ are decoupled from each other, 
and the integration over $\hcV$ 
can be performed to yield just a constant. 
Thus we obtain $F_{smm}=F_{m}$. 
Here we make a remark on the decoupling of $\hcV$. 
In the supermatrix model (\ref{supermatrix})
with $\gamma, \gamma^*\neq 0$, the
holomorphic part of the free energy $F_{smm}$ has no UV divergence,
even if we turn off the kinetic term. 
And once we do so, 
it is evident that $\hcV$ is decoupled from $\hPhi$. 
On the other hand, in the $\gamma, \gamma^*\rightarrow 0$ limit, 
we have (\ref{bncaction}) or (\ref{breduced}), 
in which the holomorphic part of the free energy becomes
UV divergent if we drop the kinetic term,
and we should introduce a regularization if we want to do so.
In other words, the kinetic term plays the role of the regularization.
And in general it is possible that a regularization induces a 
coupling between $\hPhi$ and $\hcV$,
which is universal in the sense that it does not depend on the detail
of the regularization scheme. 
We can see that this is indeed the case in (\ref{derivation}), 
where the operators that consist of $\hPhi$ are affected 
by the Konishi anomaly (\ref{anomaly}) in the $\gamma\rightarrow 0$ limit. 
In fact, as shown in (\ref{delta^2}), 
the left-hand side of (\ref{anomaly}) is finite when $\gamma\neq 0$. 
However in the $\gamma\rightarrow 0$ limit it needs some regularization 
which is the origin of the noncommutative Konishi anomaly 
(\ref{anomaly}) as we show in appendix \ref{ncKonishi}. 
This is also the case when we consider correlation functions. 
When we compute correlation functions of holomorphic quantities 
such as the resolvent (\ref{resolvent}) in the supermatrix model 
(\ref{supermatrix}), it is sufficient to consider 
the simplified supermatrix model (\ref{onesupermm}). 
However, when we take $\gamma\rightarrow 0$ limit, 
we should take account of the Konishi anomaly in
(\ref{bncaction}) and (\ref{breduced}). In fact, 
eq.(\ref{correlatorrel}) prescribes how it appears 
in the correlation function in the $\gamma\rightarrow 0$ limit 
in these theories. 
  
It immediately follows from (\ref{mmfreeE}) and (\ref{supermmfreeE}) 
that 
\be
\frac{\partial F_{m}}{\partial g_k}
=\frac{\partial F_{smm}}{\partial g_k} 
=\frac{1}{k+1}\mmfacinv\vev{\Strc\hPhi^{k+1}}. 
\label{dermmfreeE}
\ee
As shown in section \ref{SD}, 
the prepotential $\cF$ in the original gauge theory (\ref{original}) 
satisfies 
\be 
\frac{\partial \cF|_{\psi=0}}{\partial g_k} 
=\frac{1}{k+1}\coeff\vev{\tr\left(W^{\al}W_{\al}\Phi^{k+1}\right)}.
\label{derprepot}
\ee
Because in (\ref{conclusion}) we have shown directly the equivalence 
between the generating functions of (\ref{dermmfreeE}) 
and (\ref{derprepot}), we find 
\be
\frac{\partial F_{m}}{\partial g_k}
=\frac{\partial \cF|_{\psi=0}}{\partial g_k}.
\label{freeErel}
\ee
More precisely, 
${\cal F}$ is a function of $g_k$ and $S_i$ where $S_i$ 
is defined by 
\be
S_i=\frac{1}{2\pi i}\oint_{C_i}dz
    \coeff\vev{\tr\frac{W^{\al}W_{\al}}{z-\Phi}}.
\label{glueball}
\ee
{}From (\ref{conclusion}) we find that this quantity is expressed 
by the matrix model as 
\be
S_i=\frac{1}{2\pi i}\oint_{C_i}dz
    \mmfacinv\vev{\Strc\left(\resol\right)}=\frac{g_m \hN_i}{\hN}, 
\label{matrixglueball} 
\ee 
where $\hN_i$ is the number of eigenvalues of $\hPhi$ 
near the $i^{th}$ critical point. Note that in our supermatrix model, 
$\hN_i$ can take negative values,
on which we will make some comments in the next subsection. 
We emphasize that we have shown (\ref{freeErel}) as an identity. 
${\cal F}$ and $F_{m}$ 
are the same quantity. The only difference is the way 
they are represented.

In addition to this correspondence, 
we have a rather unconventional relation.
Because we have derived 
$\lim_{\gamma\rightarrow 0}S_{smm}=S_{NC}$, 
we can obtain the effective potential, or free energy 
of the noncommutative gauge theory (\ref{bncaction}) 
directly from the free energy $F_{smm}=F_{m}$ of the matrix model 
(\ref{supermatrix}) by taking the $\gamma\rightarrow 0$ limit. 
Moreover, the effective potential of (\ref{bncaction}) is independent 
of the bosonic noncommutativity $C^{\mn}$ as shown 
at the beginning of this section. It is hence the same 
as that of the commutative theory (\ref{original}). 
Therefore we obtain the following relation between
the effective potential $W_{eff}$ in the original gauge theory 
(\ref{original}) and the free energy $F_m$ of 
the simplified supermatrix model (\ref{onesupermm}):
\be 
\exp\left(-\cint~W_{eff}\right)
=\exp\left(-\frac{\hN^2}{g_m^2} F_m\right). 
\ee 
Returning to the original noncommutativities (\ref{gs}), we find  
\be
\cint~W_{eff}=\left(\invvol\right)^2 F_m, 
\label{mmfreeEandWeff} 
\ee 
which seems different from the claim of the Dijkgraaf-Vafa theory
\be
{\cal F}=F_m.
\label{FisF}
\ee
Somehow the naive use of the mapping rule gives not (\ref{FisF}) but
(\ref{mmfreeEandWeff}). 
This suggests that
$W_{eff}$ is related to ${\cal F}$ in an unconventional way 
through $F_m$. 
It would be interesting to clarify the meaning of this relation.

\subsection{Supermatrix versus bosonic matrix} 
\label{meaningofsupertr} 
In this subsection we discuss how the supermatrix model 
we have obtained (\ref{onesupermm}) is reconciled with 
the Dijkgraaf-Vafa theory, 
where the ordinary Hermitian matrix model is considered. 

Let us start with a general discussion on supermatrix. 
A Hermitian supermatrix $\hPhi$ is defined to have the following form: 
\be 
\hPhi=\left(\begin{array}{cc}
              B_1 & F_1 \\
              F_1^{\dagger} & B_2 
            \end{array} 
      \right),
\label{phi}      
\ee 
where $B_1$ and $B_2$ are $n\times n$ and $m\times m$ 
Hermitian matrices with Grassmann even entries, respectively, 
and $F_1$ is an $n\times m$ complex matrix with Grassmann odd entries. 
And the supertrace is defined by 
\be
\Str(\hPhi)=\Tr~B_1-\Tr~B_2.
\ee 
We then consider the Hermitian supermatrix model given by 
\be
S=\frac{\hN}{g_m}\Str(W(\hPhi)),
\label{onematrix}
\ee 
where $\hN=n-m$. Using the $U(n|m)$ symmetry, we can diagonalize 
$\hPhi$ as 
\be
\hPhi=U^{\dagger}
\left(\begin{array}{cccc|cccc}
              \lambda_1 & & & & & & & \\
              & \lambda_2 & & & & & & \\ 
              &  & \ddots   & & & & & \\
              & & & \lambda_{n} & & & & \\ \cline{1-8} 
              & & & &    \mu_1  & & & \\
              & & & & &   \mu_2   & & \\
              & & & & & &  \ddots   & \\
              & & & & & & & \mu_{m} \\
      \end{array}  
\right)U.
\label{diag}
\ee             
Then by rewriting (\ref{onematrix}) in terms of the eigenvalues 
$\lambda_1,...,\lambda_{n},\mu_1,...,\mu_{m}$, we obtain 
the effective action for eigenvalues
\be
S_{eff}=\frac{\hN}{g_m}\sum_{i=1}^{n}W(\lambda_i)
       -\frac{\hN}{g_m}\sum_{j=1}^{m}W(\mu_j)
       -\sum_{i<j}\log(\lambda_i-\lambda_j)^2
       -\sum_{i<j}\log(\mu_i-\mu_j)^2
       +\sum_{i,j}\log(\lambda_i-\mu_j)^2.
\ee
Note that the sign of the second and last terms are opposite 
to the ordinary Hermitian one-matrix model. 
The former is due to the supertrace, while the latter due to 
fermionic measures. Using the 
eigenvalue densities for $\lambda_i$ and $\mu_j$ 
\be 
\rho(\lambda)=\frac{1}{n}\sum_{i=1}^{n}\delta(\lambda-\lambda_i),~~~
\eta(\mu)=\frac{1}{m}\sum_{j=1}^{m}\delta(\mu-\mu_j),
\label{density}
\ee
we can rewrite this action as 
\bea
S_{eff} & = & \frac{n\hN}{g_m}\int d\lambda~\rho(\lambda)W(\lambda)
             -\frac{m\hN}{g_m}\int d\mu~ \eta(\mu)W(\mu) \nono \\
        & - & \frac{n^2}{2}\int d\lambda d\lambda'~
              \rho(\lambda)\rho(\lambda')
              \log(\lambda-\lambda')^2 \nono \\
        & - & \frac{m^2}{2}\int d\mu d\mu'~
              \eta(\mu)\eta(\mu')\log(\mu-\mu')^2 \nono \\
        & + & nm\int d\lambda d\mu~
              \rho(\lambda)\eta(\mu)\log(\lambda-\mu)^2.
\label{Seff}
\eea
If we introduce the `total' eigenvalue density defined by 
\be 
\rho_0(\lambda)=\frac{n}{\hN}\rho(\lambda)-\frac{m}{\hN}\eta(\lambda),
\label{linearcombi}
\ee
we can further rewrite $S_{eff}$ as
\be
S_{eff}=\frac{\hN^2}{g_m}\int d\lambda~\rho_0(\lambda)W(\lambda)
       -\frac{\hN^2}{2}\int d\lambda d\lambda'~
              \rho_0(\lambda)\rho_0(\lambda')
              \log(\lambda-\lambda')^2,
\ee
which is nothing but the effective action of eigenvalues 
of the ordinary Hermitian one-matrix model with the potential $W$.
 
In this sense, we can consider the ordinary Hermitian 
matrix model (\ref{hmm}) 
instead of (\ref{onesupermm}), which agrees with the 
Dijkgraaf-Vafa theory. However, in the Dijkgraaf-Vafa theory 
one should formally consider the eigenvalues which lie 
around the top of the potential. From the point of view 
of the ordinary matrix model, this is nothing but introducing 
a `negative density' of eigenvalues, which seems unnatural,
although it is formally a solution of
the Schwinger-Dyson equations.
On the other hand, this problem does not exist in the supermatrix 
model (\ref{onematrix}). 
Namely, suppose that eigenvalues around a bottom of 
the potential are regarded as those of $B_1$ 
($\lambda_i$ in (\ref{diag})), while eigenvalues 
around a top of the potential as those of $B_2$ 
($\mu_j$-type in (\ref{diag})). Then, due to the property of 
the supertrace, the eigenvalue density for the latter naturally 
appears in $S_{eff}$ with negative sign as we have seen 
in (\ref{Seff}). 
This corresponds to introducing a density with indefinite sign 
from the viewpoint of the ordinary matrix model 
as in (\ref{linearcombi}). Note here that in the supermatrix model 
the eigenvalue density $\eta(\lambda)$ itself introduced 
in (\ref{density}) 
is a positive, well-defined function. This tempts us to conclude 
that the glueball superpotential in the original gauge theory 
is described by the one-supermatrix model (\ref{onesupermm}) 
instead of (\ref{hmm}) in a rigorous sense. 

However, our supermatrix model (\ref{onesupermm}) seems to have
the following difficulty.
If we represent the noncommutative fermionic space 
in terms of Pauli matrices as in (\ref{standard}), 
the first and the second block of $\hPhi$ should have the same size,
that is, $n=m$ in (\ref{phi}).
Then from (\ref{matrixglueball}) we find that 
only the restricted domain where $\sum_iS_i=0$ 
can be described in this case. 
This drawback might originate 
from the too simple choice of the fermionic noncommutativity 
(\ref{fnc}) or the representation (\ref{standard}). 
Another possibility is that some of the eigenvalues might be 
considered to lie at infinity. 
Note that in the Dijkgraaf-Vafa theory, the extrema of the potential 
at infinity play no role if no eigenvalues lie around them.  
In this sense, there is indeed an ambiguity in the limiting procedure 
of the potential in the corresponding matrix model. 
It might be possible that 
if we take account of the kinetic term and the other terms, 
we can fix this ambiguity, and some eigenvalues are considered to 
be around the extrema at infinity. 
If this is the case, we can realize an arbitrary distribution 
of eigenvalues as a subset of the total distribution 
even if the total $S_i$ satisfies $\sum_i S_i=0$. 
In any case, it would be important to examine how we should generalize 
our supermatrix model so that it can describe more 
generic distributions of eigenvalues. We believe that 
the supermatrix model has a definite meaning, 
because it naturally arises in the mapping from the gauge theory 
to the matrix model.

\section{Discussions}
\label{discussion}
\setcounter{equation}{0}
Although we have understood essential part of the Dijkgraaf-Vafa theory 
in terms of the large-$N$ reduced model, some issues still remain 
to be clarified. 

In the Dijkgraaf-Vafa theory, the prepotential plays an 
important role in constructing the effective potential. 
In the context of the field theory, it can be understood as a result of 
the decoupling of the overall $U(1)$ part. 
However, from the point of view of the large-$N$ reduced model, 
it seems difficult to separate the $U(1)$ part and find  matrix variables 
that correspond to such fields as 
\be
w_{\al}=\frac{1}{8\pi}\tr~W_{\al}.
\ee 
In this sense, in the matrix model, the symmetry 
$W_{\al}\mapsto W_{\al}-8\pi\psi_{\al}$ can not be expressed manifestly. 
In fact, in Dijkgraaf-Vafa theory, it is conjectured that 
\be
\cF=F_m, 
\label{DVclaim}
\ee
which we have not yet shown in the matrix model context. 
Although we can prove it for the $g_k$ dependent part (\ref{freeErel}),
the reason for the full coincidence is still unclear, and   
our naive argument gives (\ref{mmfreeEandWeff}) 
instead of (\ref{DVclaim}). 
It would be an interesting problem to see how 
these structures of the ${\cal N}=2$ supersymmetry are hidden 
in the reduced model.  

As we commented at the end of subsection \ref{meaningofsupertr}, 
the apparent drawback of our supermatrix model is that it cannot 
describe arbitrary eigenvalue distributions. It is natural to expect 
that if we consider a more general noncommutative superspace
we will have a supermatrix model in which the first and the second 
blocks have different sizes. 
It would be important to deepen our understanding 
of the gauge theory on a noncommutative superspace. 

As for a generalization of our model, several directions can be 
anticipated; inclusion of a matter in the fundamental representation 
\cite{KKM}, other gauge groups, higher supersymmetries, and so on.  
It is expected that such generalizations help us to understand 
a generic structure of gauge theories 
on the noncommutative superspace, 
or supersymmetric twisted reduced models. 

In light of the ordinary twisted reduced model discussed in 
\ref{basicncgauge}, our supermatrix model (\ref{supermatrix}) 
is still unsatisfactory, because it has a dependence on the 
fermionic background $\ad\hpi_{\al}$. In order to make 
our model background-independent, it is necessary 
to introduce a gauge field associated with $\hpi_{\al}$. 
It would clarify the meaning of the fermionic noncommutativity 
$\gamma$ as a regularization, and of the Konishi anomaly 
on the noncommutative space \cite{KKM}. It would be also a clue to 
resolve the problem on the supertrace mentioned above.

\begin{center} \begin{large}
Acknowledgments
\end{large} \end{center}
We would like to thank T. Azuma, N. Ishibashi, S. Kawamoto and T. Yokono 
for useful discussions. We are grateful to R. Dijkgraaf and 
C. Vafa for informing us of their paper \cite{DV2} in which 
the relation between the Dijkgraaf-Vafa theory and the supermatrix
model is also pointed out. 
The work of T.K. was supported in part by JSPS Research Fellowships 
for Young Scientists.

\appendix
\section{Noncommutative Konishi anomaly}
\label{ncKonishi}
\setcounter{equation}{0}
In this appendix we derive the noncommutative Konishi anomaly 
(\ref{anomaly}) on the bosonic noncommutative space. 

We consider the noncommutative gauge theory (\ref{bncaction}). 
As explained in section \ref{SD}, the Konishi anomaly can be 
regarded as 
\be
\frac{\delta\Phi^i(x,\th)}{\delta\Phi^j(x',\th')}
=\delta^{i}_{j}\delta^4(x-x')\delta^2(\th-\th')
=\delta^{i}_{j}\vev{x,\th|x',\th'},
\ee
in the limit $x'\rightarrow x$ and $\th'\rightarrow \th$. 
We should evaluate this in a gauge invariant way, and 
in order to do this, we use the covariant Laplacian 
given by 
\be
\Box_{cov}=\frac{1}{16}\bD^2e^{-V}*D^2e^{V}*,
\label{Laplacian}
\ee
where $V$ is the vector superfield. 
It is easy to check that 
(\ref{Laplacian}) indeed transforms covariantly under the 
gauge transformation. In the $V\rightarrow 0$ limit, it becomes 
the ordinary Laplacian. (\ref{Laplacian}) can be also derived 
by adding the mass term of the antichiral superfield 
$\bar{m}/2\,\tr\,\bPhi*\bPhi$ to (\ref{bncaction}) 
and performing the Gaussian integration 
with respect to $\bPhi$ \cite{DGLVZ}. We evaluate 
$\vev{x,\th|x,\th}$ by the heat kernel method as follows:
\bea
\vev{x,\th|x,\th} 
 & = & \lim_{\tau\rightarrow 0}\int d^4k \kapint
       \langle x,\th|\exp_*(\tau\Box_{cov})|k,\kappa
       \rangle\langle k,\kappa|x,\th\rangle \nono \\  
 & = & \lim_{\tau\rightarrow 0}\kint 4\kapint~
       \left(\exp_*(\tau\Box_{cov})e_*^{ikx+i\th\kappa}\right)
                                   e_*^{-ikx-i\th\kappa}, \nono \\ 
 & = & \lim_{\tau\rightarrow 0}\kint 4\kapint 
       \exp_*\frac{\tau}{16}
       \left( -16k^2-\kappa^2\bD^2-8i\kappa W \right. \nono \\
 &   & \left. -4ik_{\mu}\kappa\sigma^{\mu}\bth\bD^2 
              +\bD^2e^{-V}*D^2e^V+16k_{\mu}W\sigma^{\mu}\bth
      \right).
\eea
Next we expand the exponential. First we note that 
if we use $-\kappa^2\bD^2$ or $-4ik_{\mu}\kappa\sigma^{\mu}\bth\bD^2$ 
in one of the factors in the expansion, it vanishes because 
at least one $\bD$ acts on the other factors which are chiral. 
Thus we can drop these terms in the exponential. 
Due to the integration with respect to $\kappa$, it is sufficient 
to consider the terms which contain two $\kappa$'s in the expansion 
of the exponential. However, it is easy to see 
that if such terms contain $k$, they yield positive power of $\tau$ 
after the integration with respect to $k$ and hence vanish 
in the $\tau\rightarrow 0$ limit. 
Therefore, a nonvanishing contribution comes only from 
\be
\lim_{\tau\rightarrow 0}\kint 4\kapint~
e^{-\tau k^2}\frac{\tau^2}{2}\left(-\frac{i}{2}\kappa W\right)^2_* 
=\coeff W^{\al}*W_{\al}. 
\ee 
Obviously this result is valid also in the commutative limit 
$C^{\mn}\rightarrow 0$.


\newpage

\begin{thebibliography}{99}
\bibitem{Wilson}
K.G. Wilson and J.B. Kogut,
``THE RENORMALIZATION GROUP AND THE EPSILON EXPANSION,'' 
Phys. Rept. {\bf 12} (1974) 75. 

\bibitem{EK}
T. Eguchi and H. Kawai,
``Reduction Of Dynamical Degrees Of Freedom In The Large N Gauge Theory,'' 
Phys. Rev. Lett. {\bf 48} (1982) 1063; \\
G. Parisi, ``A Simple Expression For Planar Field Theories,'' 
Phys. Lett. {\bf B112} (1982) 463; \\
D.J. Gross and Y. Kitazawa, 
``A Quenched Momentum Prescription For Large N Theories,'' 
Nucl. Phys. {\bf B206} (1982) 440; \\ 
G. Bhanot, U.M. Heller and H. Neuberger, 
``The Quenched Eguchi-Kawai Model,'' 
Phys. Lett. {\bf B113} (1982) 47; \\
S.R. Das and S.R. Wadia,
``Translation Invariance And A Reduced Model 
For Summing Planar Diagrams In QCD,''
Phys. Lett. {\bf B117} (1982) 228.     

\bibitem{GAO}
A. Gonzales-Arroyo and M. Okawa, 
``Twisted-Eguchi-Kawai Model: A Reduced Model For Large-N 
Lattice Gauge Theory,'' 
Phys. Rev. {\bf D27} (1983) 2397.

\bibitem{BFSS}
T. Banks, W. Fischler, S.H. Shenker and L. Susskind,
``M theory as a matrix model: a conjecture,''
Phys. Rev. {\bf D55} (1997) 5112, hep-th/9610043. 

\bibitem{IKKT} 
N. Ishibashi, H. Kawai, Y. Kitazawa and A. Tsuchiya,
``A large N reduced model as superstring,''
Nucl. Phys. {\bf B498} (1997) 467, hep-th/9612115.

\bibitem{IMFA}
J. Nishimura and F. Sugino,
``Dynamical generation of four-dimensional space-time 
in the IIB matrix model,''
JHEP {\bf 0205} (2002) 001, hep-th/0111102; \\
H. Kawai, S. Kawamoto, T. Kuroki, T. Matsuo and S. Shinohara, 
``Mean field approximation of IIB matrix model and emergence 
of four-dimensional space-time,''
Nucl. Phys. {\bf B647} (2002) 153, hep-th/0204240; \\
J. Nishimura, T. Okubo and F. Sugino,
``Convergent Gaussian expansion method: demonstration 
in reduced Yang-Mills integrals,''
JHEP {\bf 0210} (2002) 043, hep-th/0205253; \\  
H. Kawai, S. Kawamoto, T. Kuroki and S. Shinohara,
``Improved perturbation theory and four-dimensional space-time 
in IIB matrix model,''
Prog. Theor. Phys. {\bf 109} (2003) 115, hep-th/0211272. 

\bibitem{DV}
R. Dijkgraaf and C. Vafa, 
``A perturbative window into non-perturbative physics,'' 
hep-th/0208048.

\bibitem{Vafa}
M. Bershadsky, S. Cecotti, H. Ooguri and C.Vafa, 
``Kodaira-Spencer theory of gravity and exact results
for quantum string amplitudes,'' 
Commun. Math. Phys. {\bf165}(1994) 311 (1994), hep-th/9309140; \\ 
F. Cachazo, K.A. Intriligator and C. Vafa, 
``A large $N$ duality via a geometric transition,'' 
Nucl. Phys. {\bf B603} (2001) 3, hep-th/0103067; \\
F. Cachazo and C. Vafa, 
``${\cal N} = 1$ and ${\cal N} = 2$ geometry from fluxes,'' 
hep-th/0206017; \\
R. Dijkgraaf and C. Vafa, 
``Matrix models, topological strings, 
and supersymmetric gauge theories,'' 
Nucl. Phys. {\bf B644} (2002) 3, hep-th/0206255; \\ 
R. Dijkgraaf and C. Vafa, 
``On geometry and matrix models,'' 
Nucl. Phys. {\bf B644} (2002) 21, hep-th/0207106.

\bibitem{DGLVZ}
R. Dijkgraaf, M.T. Grisaru, C.S. Lam, C. Vafa and D. Zanon, 
``Perturbative computation of glueball superpotentials,'' 
hep-th/0211017.  

\bibitem{CDSW}
F. Cachazo, M.R. Douglas, N. Seiberg and Edward Witten, 
``Chiral rings and anomalies in supersymmetric gauge theory,'' 
JHEP {\bf 0212} (2002) 071, hep-th/0211170.

\bibitem{NCYM}
H. Aoki, N. Ishibashi, S. Iso, H. Kawai, Y. Kitazawa and T. Tada, 
``Noncommutative Yang-Mills in IIB Matrix Model,'' 
Nucl. Phys. {\bf B565} (2000) 176, hep-th/9908141; \\ 
N. Ishibashi, S. Iso, H. Kawai and Y. Kitazawa,
``Wilson loops in noncommutative Yang-Mills,'' 
Nucl. Phys. {\bf B573} (2000) 573, hep-th/9910004. 

\bibitem{Konishi}
K. Konishi, 
``Anomalous Supersymmetry Transformation OF Some Composite 
Operators IN Sqcd,'' Phys. Lett. {\bf B135} (1984) 439; \\
K. Konishi and K. Shizuya, 
``Functional Integral Approach To Chiral 
Anomalies In Supersymmetric Gauge Theories,'' 
Nuovo Cim. {\bf A90} (1985) 111.  

\bibitem{SW}
A. Connes, M.R. Douglas and A. Schwarz, 
``Noncommutative geometry and matrix theory: compactification on tori, '' 
JHEP {\bf 9802} (1998) 003, hep-th/9711162; \\
M.R. Douglas and C.M. Hull, 
``D-branes and the noncommutative torus,'' 
JHEP {\bf 9802} (1998) 008, hep-th/9711165; \\
N. Seiberg and E. Witten, 
``String theory and noncommutative geometry,''
JHEP {\bf 9909} (1999) 032, hep-th/9908142.   

\bibitem{IRUV}
S. Minwalla, M. Van Raamsdonk and N. Seiberg, 
``Noncommutative perturbative dynamics,''
JHEP {\bf 0002} (2000) 020, hep-th/9912072; \\
M. Van Raamsdonk and N. Seiberg, 
``Comments on noncommutative perturbative dynamics,''
JHEP {\bf 0003} (2000) 035, hep-th/0002186.  

\bibitem{DV2}
R. Dijkgraaf and C. Vafa, 
``${\cal N}=1$ Supersymmetry, deconstruction, and bosonic gauge
theories,'' 
hep-th/0302011.

\bibitem{KKM}
H. Kawai, T. Kuroki and T. Morita, work in progress.

\end{thebibliography}
\end{document}